\documentclass[12pt,preprint]{aastex}

\usepackage{amsmath}
\usepackage{amssymb}
\newcommand{\bfm}[1]{\mbox{\boldmath{$#1$}}}
\newcommand{\ds}{\displaystyle}
\usepackage{stmaryrd}

\usepackage{appendix}
\usepackage{color}

\usepackage{todonotes}

\shorttitle{Chaos in Navigation Satellite Orbits}
\shortauthors{Rosengren et al.}

\begin{document}

\title{From order to chaos in Earth satellite orbits}        
\author{Ioannis Gkolias,\altaffilmark{1}
	J\'{e}r\^{o}me Daquin,\altaffilmark{2,3} 
	Fabien Gachet\altaffilmark{1} \&
	Aaron J. Rosengren\altaffilmark{4,5}}

\affil{\altaffilmark{1}Department of Mathematics, University of Rome Tor Vergata,
	I-00133 Rome, Italy}
\affil{\altaffilmark{2}IMCCE/Observatoire de Paris, Universit\'{e} Lille1,
	F-59000 Lille, France}	
\affil{\altaffilmark{3}SPACE Research Centre, School of Mathematical and Geospatial Sciences, 
	Royal Melbourne Institute of Technology (RMIT) University,
	Melbourne 3001, Australia}
\affil{\altaffilmark{4}IFAC-CNR,
	50019 Sesto Fiorentino, Florence, Italy}
\affil{\altaffilmark{5}Department of Physics, Aristotle University of Thessaloniki, 
	54124 Thessalon'ki, Greece}
	
\email{gkolias@mat.uniroma2.it}

\begin{abstract}

We consider Earth satellite orbits in the range of semi-major axes where the perturbing effects of Earth's oblateness and lunisolar gravity are of comparable order. This range covers the medium-Earth orbits (MEO) of the Global Navigation Satellite Systems and the geosynchronous orbits (GEO) of the communication satellites. We recall a secular and quadrupolar model, based on the Milankovitch vector formulation of perturbation theory, which governs the long-term orbital evolution subject to the predominant gravitational interactions. We study the global dynamics of this two-and-a-half degrees-of-freedom Hamiltonian system by means of the fast Lyapunov indicator (FLI), used in a statistical sense. Specifically, we characterize the degree of chaoticity of the action space using angle-averaged normalized FLI maps, thereby overcoming the angle dependencies of the conventional stability maps. Emphasis is placed upon the phase-space structures near secular resonances, which are of first importance to the space debris community. We confirm and quantify the transition from order to chaos in MEO, stemming from the critical inclinations, and find that highly inclined GEO orbits are particularly unstable. Despite their reputed normality, Earth satellite orbits can possess an extraordinarily rich spectrum of dynamical behaviors, and, from a mathematical perspective, have all the complications that make them very interesting candidates for testing the modern tools of chaos theory.

\end{abstract}


\section{INTRODUCTION}

With sizes ranging from chunks of rock less than a kilometer across to spherical objects larger than Mercury, the natural planetary satellites in the Solar System present a great variety of perturbation problems \citep{sP99}, with the main disturbing forces comprising the oblateness of the primary and, in the case of Saturn's satellites, the attraction of the rings, the attraction by the Sun, and the attraction by other satellites revolving around the same planet. In the absence of mutual satellite mean-motion interactions \citep[qq.v.][]{nC15,rF15}, the orbital dynamics of most satellites is primarily governed by secular gravitational effects \citep{gS28,rG75,rM89,sT09}. The motion of a satellite in the field of an oblate spheroid disturbed by the gravitational attraction of a distant perturber was treated as early as the 1800s in Laplace's most splendid `M\'ecanique C\'eleste' \citep{dBgC61}. This subject was reinvigorated with the advent of artificial satellites of the Earth (1957) and the Moon (1966). Unlike the natural systems, where generally the orbit perturbations of one type determine the character of the motion (the outstanding exception being Saturn's moon, Iapetus), many artificial satellites reside in the region where the secular perturbing effects due to Earth's oblateness and lunisolar forces are of comparable order \citep{mH81,vK94,tE97,cZ15}. Although the mean eccentricities and inclinations of the orbits of the disturbing bodies are not unduly large ($0.0549$ and $5^\circ9^\prime$, respectively, for the Moon), artificial satellites may be launched in any eccentric orbit ($0 < e < 1$) with inclinations lying anywhere between $0^\circ$ and $180^\circ$. Here, the time-honored methods developed over the last three centuries for classical Solar System dynamics are not completely applicable. 

Today, these artificial satellites have had such thorough attention accorded to their study that the outstanding difficulties are thought to be solely in the final refinements \citep[e.g., tracing the most minute consequences of yet another tesseral harmonic, to paraphrase][]{sB01}. Despite nearly sixty years of space activities, however, we cannot yet construct a complete, logically ordered picture of the global dynamics of gravitationally dominated satellite orbits around the Earth. Indeed, there is no complete general solution for the long-term behavior of such satellites even in the averaged problem under the quadrupolar approximation. That is, when the disturbing function, consisting of the dominate oblateness term in the geopotential ($J_2$) together with the lowest-order term (second harmonic) in the Legendre expansion of both the lunar and solar potentials (i.e., Hill's approximation), is averaged over the periods of the satellite and the disturbing bodies \citep[q.v.][]{rAgC64}. Under the further approximation that the lunar orbit lies in the ecliptic, reducing the averaged system to two degrees of freedom, \citet{mLmY74} indicate the known integrable cases and investigate the geometrical behavior of the resulting motion. Certain limiting cases, such as the critical inclination problem and the Lidov--Kozai oscillations \citep[q.v.][]{sN16}, have been studied extensively. The Lidov--Kozai mechanism can produce large-amplitude oscillations of the eccentricity and inclination on timescales long compared to the orbital periods. However, for most artificial satellites, this effect is weakened or completely suppressed by the oblateness precession \citep{bL15}, especially in regions close to the Earth \citep{sT14}. The dynamical effects of oblateness thereby enhance the stability of satellite orbits, except in small intervals in the inclination centered on resonances between the two secular precession frequencies \citep{mV74}. These narrow instability zones, in which eccentricity grows exponentially \citep{bKsD11}, are found near the critical inclinations, $46\fdg 4$, $56\fdg 1$, $63\fdg 4$, $69\fdg 0$, $73\fdg 2$, and their respective supplementary angles for retrograde orbits.

The problem becomes increasingly more difficult when the complexities of the Earth--Moon--Sun system are taken into account, giving rise to additional secular resonances involving commensurabilities amongst the slow frequencies of orbital precession of the bodies involved. While the existence of some of these resonances has been known since the early Sputnik era \citep{pM59}, the implications of their dynamical effects are still being fully probed and understood \citep{aR15a,aR15b,jD16a,jD16b,aC16,aCetal}. In \citet{jD16a}, it was shown that the devious network of lunisolar secular resonances that permeate the phase space of the highly inclined navigation satellites can interact to produce chaotic and diffusive motions. Using the fast Lyapunov indicator (FLI), they presented dynamical stability maps that revealed a transition from a mostly stable region at three Earth radii, where regular orbits dominate, to a resonance overlapping and chaotically connected one at five Earth radii. The very limited region within which \citet{jD16a} were able to survey the long-term motions calls forth attempts to idealize the problem even further, in order to make a more efficient numerical treatment possible without at the same time deviating too drastically from the actual conditions prevailing in the system. The fact that \citet{jD16a} have already isolated the basic physical model governing the long-term evolution of satellite orbits in the medium-Earth orbit (MEO) region, and furthermore that the occurrence and nature of the secular resonances driving these dynamics depend chiefly on the frequencies of nodal and apsidal precession and the rate of regression of the lunar nodes \citep{aR15a}, suggests the use of the simple and direct model of quadrupole-order, secular oblateness and third-body perturbations of \citet{sT09}, suitably modified to account for the most relevant irregularity in the Moon's motion.
  
Here, we recall this doubly averaged model \citep{rAgC64,sT09}, which is based on the Milankovitch vector formulation of perturbation theory, and show that its systematic application leads to a successful picture of the dynamical structure of the competing region of oblateness and lunisolar perturbations. The averaged equations of motion hold rigorously for all Keplerian orbits with nonzero angular momentum, and, along with their variational equations (reported herein for the first time), are given in a concise analytical vector form, which also intrinsically accounts for the Moon's perturbed motion \citep[q.v.][]{aRdS13}. The goal of this paper is to explore in more detail the transition from order to chaos in Earth satellite orbits using this vector formulation, and to extend the inclination--eccentricity phase-space study of \citet{jD16a} beyond MEO to seven Earth radii and to retrograde orbits. This work aims at casting more light on the lunisolar secular resonance problem by considering the most basic physical model that accurately reflects the true nature of the resonant interactions. In this respect, we also partially address a fundamental question left open by \citet{sT09} on the stability of general satellite orbits in quadrupolar fields. The recognition of the importance of the perturber's orbital axis precession raises many questions of interest concerning quadrupole-order secular evolution. 

\section{PROBLEM FORMULATION}
\label{sec:Milan}

The integrable Kepler problem so frequently approximates the real state of affairs that it has been for over three centuries the conceptual basis for most mathematical formulations in celestial mechanics. In a satellite's orbital motion around the Earth, the existence of the other members of the system (i.e., the Moon or the Sun) or any irregularity in the shape and density of the Earth, manifests itself mainly by rather small or slowly changing perturbations of the elements of the Keplerian ellipse. Qualitatively, the perturbed Keplerian orbit gains two additional frequencies, consisting of a precession of the orbit plane and a rotation of the major axis in the moving orbit plane (nodal and apsidal motion), which are slow relative to the orbital frequency (i.e., the mean motion). If there are no commensurability relationships involving the mean motions, the Hamiltonian can be averaged over the successive orbital phases (mean anomalies) of the system to obtain the secular equations of motion.

In the secular approximation, the semi-major axis $a$ becomes a constant of motion and the problem reduces to understanding the long-term behaviors of the four remaining orbital elements, inclination $i$, eccentricity $e$, longitude of the ascending node $\Omega$, and argument of periapsis $\omega$, at a given semi-major axis. The orbit can be parameterized by two vectors, well defined for all bound orbits: the dimensionless angular momentum vector ${\bfm h} = \sqrt{1 - e^2} \hat{\bfm h}$, pointing perpendicular to the orbital plane, and the eccentricity vector ${\bfm e} = e \hat{\bfm e}$, defining the orientation of the major axis in the orbital plane \citep[qq.v.][]{pM61,rAgC64,sT09,bKsD11,dS12,aRdS13,sT14,bL15}. These vectorial elements are free of singularities associated with zero eccentricity, vanishing line of nodes, polar motion, and straight-line degeneracy. We can describe the orientation-defining integrals of the two-body problem, $\hat{\bfm h}$ and $\hat{\bfm e}$, in terms of the classical orbital elements relative to an inertial frame as $\hat{\bfm h} = \sin i \sin \Omega \hat{\bfm x} - \sin i \cos \Omega \hat{\bfm y} + \cos i \hat{\bfm z}$ and $\hat{\bfm e} = (\cos \omega \cos \Omega - \cos i \sin \omega \sin \Omega) \hat{\bfm x} + (\cos \omega \sin \Omega + \cos i \sin \omega \cos \Omega) \hat{\bfm y} + \sin i \sin \omega \hat{\bfm z}$. 

\subsection{The Orbit-averaged Hamiltonian}

For satellites sufficiently far from the Earth's surface, the perturbing potential arising from the irregularities of the Earth's gravitational field can be described with reasonable accuracy by the quadrupole moment from the equatorial bulge. And taking into consideration the smallness of the ratio of the semi-major axes of the satellite and the disturbing bodies (being $\lesssim 0.1$ for the Moon and $\ll 0.1$ for the Sun, for the orbits of most satellites), the lunar and solar perturbing potentials may be approximated by quadrupolar fields (i.e., their potentials are limited to the second Legendre polynomial term, like in the Hill problem). In this hierarchical configuration, the secular orbital evolution of the satellite is determined by a double time-average over the mean anomalies of the bodies involved, assuming that all of the orbital variables except the respective orbital phases remain constant during the averaging procedures. A reasonably accurate representation of the orbital evolution should be obtained if the main oscillation of the predicted motion has a period greater than a year \citep{sB01}. 

We can write the doubly averaged potential associated with the quadrupole-order oblateness and lunisolar perturbations as $\Phi = n a^2 \Phi^\ast$, where \citep{rAgC64,sT09,sT14}
\begin{equation}
	\Phi^\ast \label{eq:potential}
	= \frac{\omega_0}{6 h^3} \left[ 1 - 3 (\hat{\bfm p} \cdot \hat{\bfm h})^2 \right] 
 	+ {\textstyle\sum\limits_{j=1,2}}\, \frac{\omega_j}{2} \left[ 5 (\hat{\bfm H}_j \cdot {\bfm e})^2 
		- (\hat{\bfm H}_j \cdot {\bfm h})^2 - 2 e^2 \right],
\end{equation}
Here $\hat{\bfm H}_j$ is a unit vector aligned with the disturbing body's orbit pole (i.e., $\hat{\bfm H}_1$ lies along the pole of the ecliptic and $\hat{\bfm H}_2$ is normal to the lunar orbit plane); $\hat{\bfm p}$ is a unit vector aligned with the Earth's spin axis; and  
\begin{equation}
	\label{eq:precession_factors}
	\omega_0 = \frac{3 n J_2 R^2}{2 a^2},\quad \omega_j = \frac{3 \mu_j}{4 n a_j^3 (1 - e_j^2)^{3/2}}.
\end{equation}
The satellite's semi-major axis $a$ is constant in the secular approximation, $n = \sqrt{\mu/a^3}$ is the mean motion, $\mu$ and $\mu_j$ are the respective gravitational parameters of the Earth and perturbing bodies, $a_j$ and $e_j$ are the orbit semi-major axis and eccentricity of the perturbing bodies, $J_2$ is the quadrupole moment of Earth's gravitational potential (the coefficient of the second zonal harmonic), and $R$ is the Earth's mean equatorial radius. 

We have restricted our analysis to quadrupolar gravitational interactions by retaining only the second harmonic in the Legendre polynomial series expansions of both the geopotential and those associated with lunisolar perturbations. It is quite fortuitous that under this treatment, the long-period lunar and solar effects, notes \citet{pM61}, depend only upon the position of the orbital planes of the disturbing bodies and are not influenced by the position of their perigees. The decisive significance of this fact on the integrability of the secular hierarchical restricted three-body problem and for the existence of the Lidov--Kozai resonance was emphasized in the early 1960s \citep[q.v.][and references therein]{sN16}. The basic assumption made here is on the satellite's semi-major axis, which naturally limits the class of orbits that can be handled. The period of oblateness precession, whose rate decreases as $a^{-7/2}$, must be appreciably longer than a year for the double-averaging procedure (and the neglect of higher-order gravity field perturbations) to be justifiable, thereby setting a lower limit to the orbital radius.\footnote{Note that the satellite's semi-major axis is not the only factor constraining apsidal and nodal precession. As far as the $J_2$ effects alone are concerned, both precessional motions increase with orbit eccentricity in the same manner (to terms of order $e^2$). Both precessional motions vary with orbit inclination to the equator, with nodal precession being greatest for equatorial orbits and vanishing for polar orbits.} For the solar potential, the effects of the third harmonic (parallactic, or octupole) term is negligible, but its contribution to the lunar perturbations becomes increasingly more important as $a$ increases, which sets an upper limit \citep{pM61}. As a result, the analysis is most useful in the region of semi-major axes between three and six Earth radii. However, the omission of the lunar parallactic term for geosynchronous orbits (GEO) ($a/a_2\sim0.11$) is not as serious as it might at first appear \citep{aRdS13}. Bearing these caveats in mind, here we will treat here orbits with semi-major axes up to seven Earth radii.   

The averaged Hamiltonian is given by
\begin{equation}
	\label{eq:hamil}
	\mathcal{H} = \mathcal{H}_\text{kep} + \Phi,
\end{equation}
where the Kepler Hamiltonian is $\mathcal{H}_\text{kep} = -\mu/2a$ depends only on the mean semi-major axis. Despite the Hamiltonian being averaged over the satellite mean anomaly, and over the lunar and apparent solar orbital motions, it is still time dependent due to the secular motion of the Moon's orbital nodes. The study of the orbital motion of Earth satellites is, therefore, intricately entangled with the perturbed motion of the Moon \citep{aR15a}. Following a line of thought laid down by Horrocks and Hill in their lunar theories, modifying the form to serve our purposes, we can detach one of the most pronounced perturbations of the Sun and amalgamate its effect with the motion of the Kepler ellipse. Accordingly, we regard the lunar orbit as unperturbed, except for a slow uniform regression of its nodes (period of $\sim18.6$ years). Of course, this differs substantially from the true motion of the Moon, yet, this simplified treatment is found to be satisfactory, as far as the perturbations of a satellite's orbit are concerned (see Appendix~\ref{sec:validate}). 

\subsection{Secular Equations of Motion and Variational System}

Our basic model forms a non-autonomous Hamiltonian two degrees-of-freedom system, depending periodically on time through the lunar nodal motion. The secular equations of motion arising from Earth's oblateness and lunisolar perturbations follow from the Lagrange equations, and can be stated compactly and elegantly in dyadic form as\footnote{The notation $\widetilde{\bfm a}$ denotes the cross-product dyadic, defined such that $\widetilde{\bfm a} \cdot {\bfm b} = {\bfm a} \cdot \widetilde{\bfm b} = {\bfm a} \times {\bfm b}$. The product of two unit vectors, ${\bfm a} {\bfm b}$, is called a dyad and, in column and row vector notation, is equivalent to the outer product (i.e., $[{\bfm a}] [{\bfm b}]^\intercal$). The identity dyadic is denoted by ${\bfm U}$ and has the general property ${\bfm U} \cdot {\bfm a} = {\bfm a} \cdot {\bfm U} = {\bfm a}$.}
\citep{rAgC64,sT09,aRdS13}
\begin{equation}
\label{eq:secular_dynamics}
{\arraycolsep=0.6em\def\arraystretch{2.3}
\begin{array}{l}
	\ds\dot{\bfm h}
	=-\frac{\omega_0}{h^5}(\hat{\bfm p} \cdot \bfm{h}){\widetilde{\hat{\bfm p}}} \cdot \bfm{h}
		-{\textstyle\sum\limits_{j=1,2}}\, \omega_j \hat{\bfm H}_j \cdot \left( 5 \bfm{e} \bfm{e}
		-\bfm{h} \bfm{h} \right) \cdot{\widetilde{\hat{\bfm H}}}_j, \\
	\ds\dot{\bfm e}
	=-\frac{\omega_0}{2h^5} \left\{ \left[ 1-\frac{5}{h^2} (\hat{\bfm p} \cdot \bfm{h})^2 \right]
		{\widetilde{\bfm h}} + 2 (\hat{\bfm p} \cdot \bfm{h}){\widetilde{\hat{\bfm p}}} \right\} \cdot \bfm{e} \\
	\hspace{23pt} \ds - {\textstyle\sum\limits_{j=1,2}}\omega_j \left[ \hat{\bfm H}_j \cdot 
		\left( 5 \bfm{e} \bfm{h} - \bfm{h} \bfm{e} \right) \cdot {\widetilde{\hat{\bfm H}}}_j
		- 2 {\widetilde{\bfm h}} \cdot \bfm{e} \right].
\end{array}}
\end{equation}
Physical solutions are restricted to those satisfying the constraints ${\bfm h} \cdot {\bfm e} = 0$ and ${\bfm h} \cdot {\bfm h} + {\bfm e} \cdot {\bfm e} = 1$. 

We are indebted to \citet{sT09} for an exhaustive treatment concerning the study of the corresponding autonomous, two degrees-of-freedom system of Equations~\ref{eq:secular_dynamics}; that is, when the axes ${\bfm H}_j$ are fixed so that $\mathcal{H}$ is independent of explicit time. In particular, they studied the properties of the classical Laplace plane \citep[qq.v.][]{dBgC61,rAgC64}, including the stability of its generating orbits and its generalization to eccentric orbits \citep[see also][for a non-vectorial treatment of the problem]{vK97}. However, their work left open the question of the dynamical structure of the general four-dimensional system, and, particularly, the regions of phase space in which chaotic orbits can be found. For higher-dimensional dynamics systems, such as that considered here, this issue can easily be explored through the numerical computation of chaos indicators \citep[q.v.][]{cF00}, which measure the asymptotic growth rate of the length of deviation vectors. Hence, to study the stability of a general orbit, we form the full first-order variational equations.

Assume a nominal trajectory ${\bfm x} (t) = \left[ {\bfm h} (t); {\bfm e} (t) \right]$; linearizing about the nominal trajectory yields the Jacobian matrix, which takes a small variation ${\bfm w}$ into a later variation relative to the nominal trajectory: 
\begin{equation}
	\dot{\bfm w} = {\bfm \Phi} \cdot {\bfm w}, 
\end{equation}
where
\begin{equation}
{\bfm \Phi} (t) = 
{\arraycolsep=0.6em\def\arraystretch{2.3}
	\left[ \begin{array}{ll} 
		 \ds \frac{\partial \dot{\bfm h}}{\partial {\bfm h}} & \ds \frac{\partial \dot{\bfm h}}{\partial {\bfm e}} \\
		  \ds \frac{\partial \dot{\bfm e}}{\partial {\bfm h}} & \ds \frac{\partial \dot{\bfm e}}{\partial {\bfm e}}
	\end{array} \right]}, 
\end{equation}
and
\begin{equation}
\label{eq:var_dynamics}
{\arraycolsep=0.6em\def\arraystretch{2.3}
\begin{array}{l}
	\ds \frac{\partial \dot{\bfm h}}{\partial {\bfm h}} 
	= \frac{\omega_0}{h^5} \left[ {\widetilde{\bfm h}} 
		\cdot \hat{\bfm p} \hat{\bfm p} \cdot \left( {\bfm U} - 5 \hat{\bfm h} \hat{\bfm h} \right) 
		- ( \hat{\bfm p} \cdot {\bfm h} ) {\widetilde{\hat{\bfm p}}} \right]
		+ {\textstyle\sum\limits_{j=1,2}} \omega_j \left[ {\widetilde{\bfm h}} \cdot \hat{\bfm H}_j \hat{\bfm H}_j
		- ( \hat{\bfm H}_j \cdot {\bfm h} ) {\widetilde{\hat{\bfm H}}}_j \right], \\
	\ds \frac{\partial \dot{\bfm h}}{\partial {\bfm e}} 
	= - {\textstyle\sum\limits_{j=1,2}} 5 \omega_j 
			\left[ {\widetilde{\bfm e}} \cdot \hat{\bfm H}_j \hat{\bfm H}_j 
			- ( \hat{\bfm H}_j \cdot {\bfm e} ) {\widetilde{\hat{\bfm H}}}_j \right], \\		
	\ds \frac{\partial \dot{\bfm e}}{\partial {\bfm h}} \nonumber
	= \frac{\omega_0}{2 h^5} {\widetilde{\bfm e}} \cdot \left\{ \left[ \left( 1 - \frac{5}{h^2} 
			(\hat{\bfm p} \cdot \bfm{h})^2 \right) {\bfm U} + 2 \hat{\bfm p} \hat{\bfm p} \right] 
			\cdot \left( {\bfm U} - 5 \hat{\bfm h} \hat{\bfm h} \right)
			+ \frac{10}{h^2} (\hat{\bfm p} \cdot {\bfm h}) 
			\left[ (\hat{\bfm p} \cdot {\bfm h}) \hat{\bfm h} \hat{\bfm h} - {\bfm h} \hat{\bfm p} \right] \right\} \\
	\ds \hspace{32pt} \left. + {\textstyle\sum\limits_{j=1,2}} \omega_j 
			\left[ {\widetilde{\bfm e}} \cdot \hat{\bfm H}_j \hat{\bfm H}_j
			+ 5 ( \hat{\bfm H}_j \cdot {\bfm e} ) {\widetilde{\hat{\bfm H}}}_j 
			- 2 {\widetilde{\bfm e}} \right] \right\}, \\
	\ds \frac{\partial \dot{\bfm e}}{\partial {\bfm e}} 
	= -\frac{\omega_0}{2 h^5} \left\{ \left[ 1-\frac{5}{h^2} (\hat{\bfm p} \cdot \bfm{h})^2 \right] 
			{\widetilde{\bfm h}} + 2 ( \hat{\bfm p} \cdot {\bfm h} ) {\widetilde{\hat{\bfm p}}} \right\} 
			- {\textstyle\sum\limits_{j=1,2}} \omega_j \left[ 5 {\widetilde{\bfm h}} 
			\cdot \hat{\bfm H}_j \hat{\bfm H}_j + ( \hat{\bfm H}_j \cdot {\bfm h} ) {\widetilde{\hat{\bfm H}}}_j 
			- 2 {\widetilde{\bfm h}} \right].
\end{array}}
\end{equation}

\section{SECULAR DYNAMICS OF MEDIUM- AND HIGH-EARTH ORBITS}

The global properties of multidimensional, nearly integrable Hamiltonian systems are determined by the relative location and size of the nonlinear resonances \citep{aM02}. Multi-frequency systems with two or more resonances, involving two or more degrees of freedom, can undergo a transition from order to dynamical chaos in certain regions of phase space, as the system parameters are varied \citep{cF00}. Regular behavior is destroyed from the growth and overlap of resonances, and, as a result, the states of the system are free to explore in an apparently random manner throughout the phase-space region of influence of the interacting resonances. This order-to-chaos transition scenario was found by \citet{jD16a} to occur in the highly inclined MEOs of the navigation satellites, brought on by lunisolar secular resonances dependent only on the frequencies of nodal and apsidal precession and the rate of regression of the Moon's nodes. As the semi-major axis recedes from Earth, scanning the MEO region, they found (both analytically and numerically) a transition from a globally stable regime at three Earth radii, where resonances are thin and well separated, to a globally unstable one at five Earth radii, in which the resonances become stronger and overlap. The lunisolar secular resonance zones form a complicated network through the inclination and eccentricity phase space \citep{tE97,aR15a,jD16a,aCetal}, and an orbit that lies in the stochastic layer can eventually diffuse (in Fick's sense) into different regions of the phase space by moving across or along the resonances \citep{jD16b}. 

\citet{jD16a}, to which we refer for more details, have given a clear picture of the structure, extent, and evolution of the chaotic regions in MEO, using FLI stability maps. Nevertheless, a systematic investigation of this complicated phenomenon over the relevant semi-major axis scales is hitherto missing. On account of the recent works that found large-scale variations in eccentricity in the highly inclined geosynchronous region \citep{vM14,cZ15}, we extend here the \citet{jD16a} phase-space stability study to seven Earth radii. Although artificial satellites are rarely launched into retrograde orbits at these geocentric distances, for completeness we also consider the retrograde regime. However, one of the principal limitations of FLI maps is their dependence on the initial phases \citep[q.v.][and references therein]{nTbN15}. That is, these stability maps characterize the dynamics in the inclination and eccentricity (`action') phase space for fixed initial angles. While the choice of the initial phases may not drastically change the general dynamical picture of the global phase space \citep{pR01,nTbN15}, it can have a significant impact on the dynamics locally, particularly inside the libration regions \citep{jD16a}. Indeed, not only does the resonant geometry generally depend on the initial phase angles, but there will also be differences in the crossings of the resonant regions \citep{pR01}. To make this more concrete, the role of the initial phases is clearer if we look at the phase portrait of a resonance. Choosing then for the maps the angles corresponding to the hyperbolic or elliptic fixed points of the resonance would mean avoiding stable islands or the opposite, with all other angular variables giving some kind of intermediate situation. The problem is that each resonance has its own specific values of angles, and, although these combinations may be limited by the number of primary resonances, there is an ambiguity in the selection when the emphasis is on the global dynamics of the system. 

While it is inherently difficult to capture the complete dynamics of a six-dimensional phase space in the plane of dimension two, \citet{nTbN15} have suggested a `random-angles' FLI map in order to obtain a better estimate of the real width of the chaotic zones associated to each resonance. In this way, the FLI maps are determined using random initial phase angles, which, while being no more computationally expensive than the traditional FLI calculations, compromises the sharpness of the pictures and hides the rich structures within the chaotic zones. To this end, we use here an `angle-averaged' normalized FLI map (Appendix~\ref{sec:FLI}), which preserves the phase-space structure and allows for an angles-independent, asymptotic characterization of the action plane. Each dynamical stability map produced in this way requires the computation of a large set of traditional FLI maps, each with a fixed random set of initial angles. Such a formidable computational task is only possible with our simplified dynamical model, on account of the double averaging and basic lunar model significantly reducing the numerical burden.

\begin{figure}[htp!]
	\centering
	\includegraphics[width=0.5\textwidth]{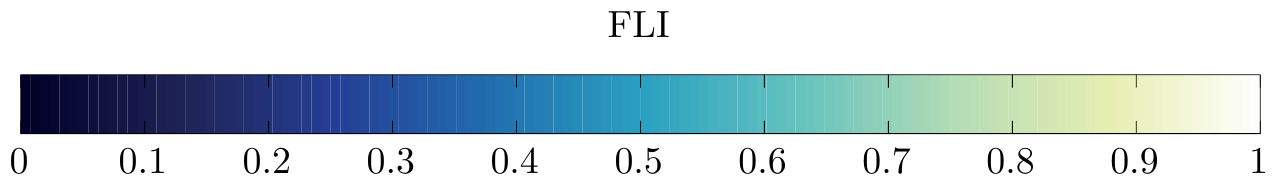}
	\includegraphics[width=0.65\textwidth]{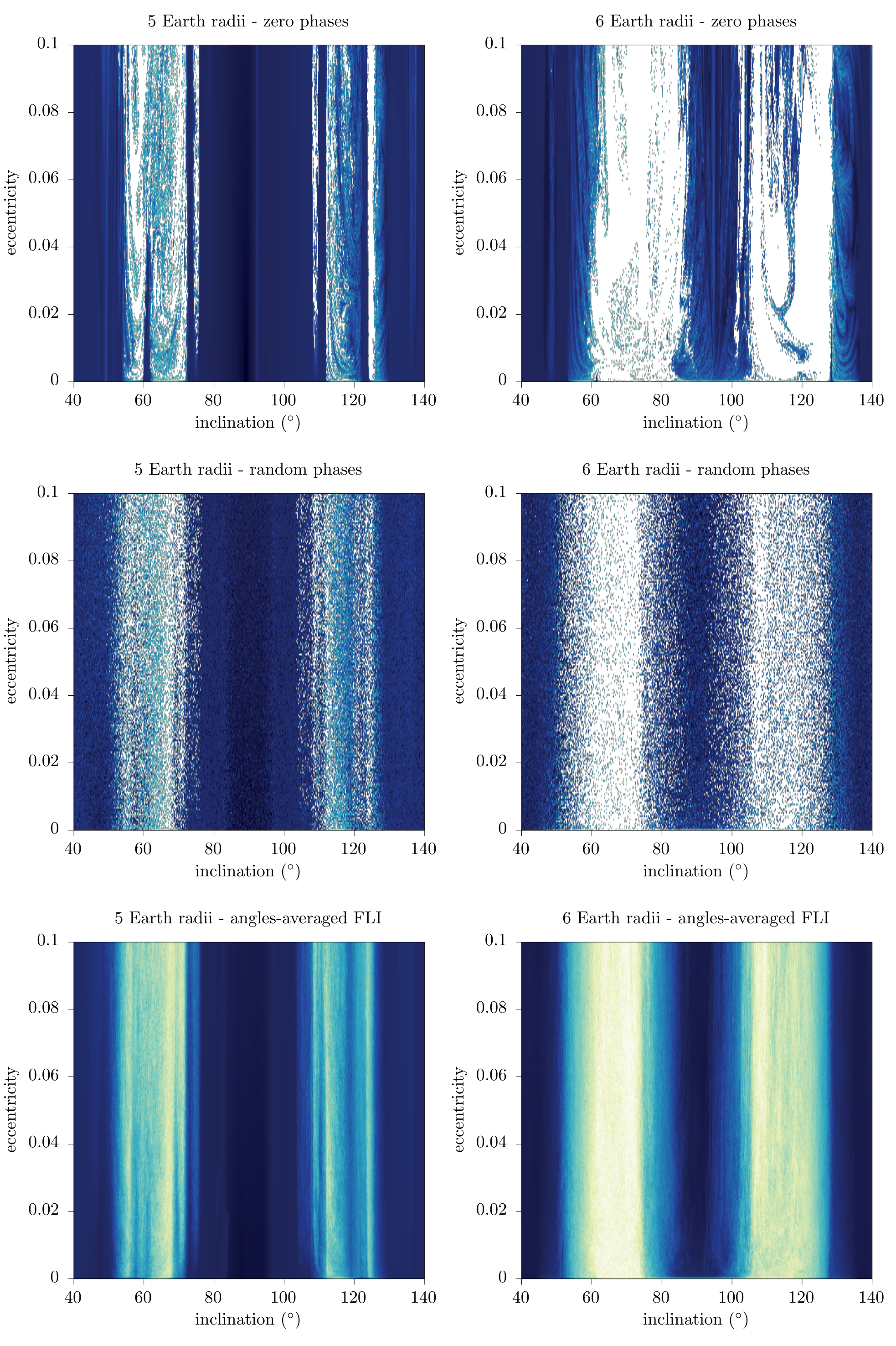}	
	\caption{Dynamical stability maps in the inclination--eccentricity phase space produced using the conventional FLI with initial phases set to zero (top), the FLI with randomly taken initial angles for each grid point (middle), and the angle-averaged FLI 	obtained from 50 distinct FLI maps (bottom). The colorbar for the FLI maps is normalized to 1.}
	\label{fig:FLI3methods}
\end{figure}

Figure~\ref{fig:FLI3methods} shows the FLI stability maps in a grid of $200 \times 500$ initial conditions of the $e-i$ plane (focusing on the critical inclinations) for semi-major axes of five and six Earth radii. In the top panel, the FLI is computed in the conventional sense with all initial angles ($\omega$, $\Omega$, and $\Omega_2$) set to zero. Each orbit is propagated up to 465 years (about 25 lunar nodal periods), and Earth re-entry orbits were declared if the perigee radius $a (1 - e)$ drops below an altitude of $120$ km (appearing white in the maps and given an FLI value of 1). The darker blue regions correspond to stable orbits, where the variations are regular, while the lighter green and yellow areas mark the chaotic zones. Note that because the secular system, Equations~\ref{eq:secular_dynamics}, is invariant under the transformation ${\bfm e} \rightarrow -{\bfm e}$ (or, equivalently, $\omega \rightarrow \omega + \pi$), the range of argument of perigee can be restricted from $[0, 2\pi]$ to $[0, \pi]$. Accordingly, the middle panel shows the FLI maps computed with orbital angles taken randomly from the intervals $\omega \in [0, \pi]$, $\Omega, \Omega_2 \in [0, 2 \pi]$, and the bottom panel gives the synthesis of 50 distinct FLI maps for each semi-major axis, with the fixed angles for these individual computations being sampled from the same intervals (each map thereby representing five million orbits).\footnote{The ensemble size of 50 for the averaged FLI maps was specified by accumulating maps until the results displayed no more sensitivity to further additions. Even doubling the number of ensembles yielded the same averaged map.} The averaged FLI captures the full extent of the chaotic zones, without muddling the stable regions within these bands, and can be effectively used to classify the degree of hyperbolicity within the action space across all angles. 

\begin{figure}[htp!]
	\centering
	\includegraphics[width=0.6\textwidth]{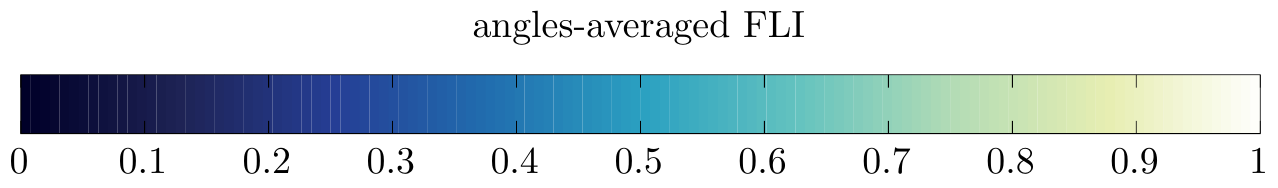}
	\includegraphics[width=\textwidth]{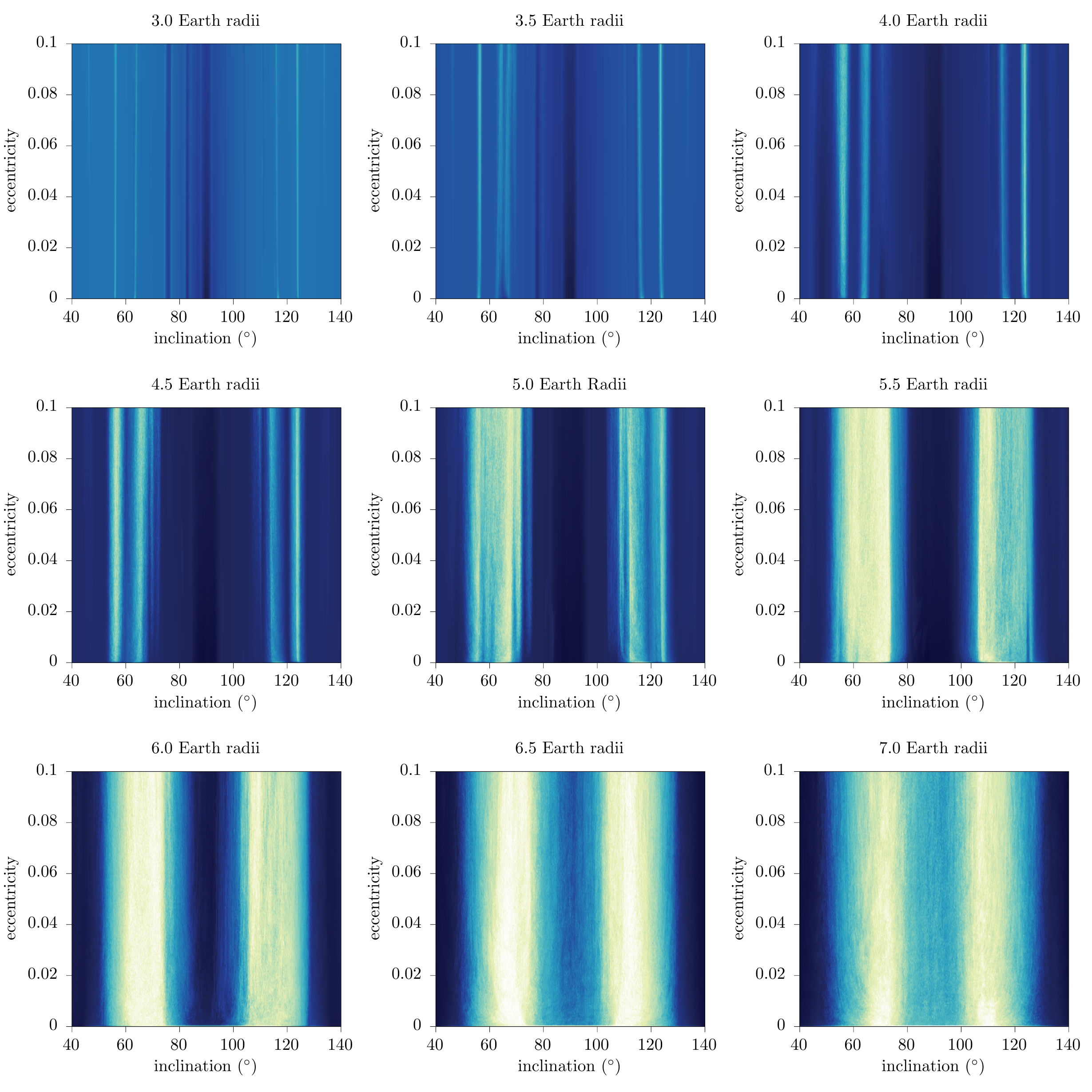}
	\caption{Dynamical stability maps in the inclination--eccentricity phase space produced using the angle-averaged normalized FLI for increasing values of the satellites' semi-major axis. The colorbar for the FLI maps is normalized to 1.}
	\label{fig:ecc_vs_inc}
\end{figure}

We carried out a corresponding numerical survey over the entire domain of validity of our model, using the averaged FLI technique, as shown in Figure~\ref{fig:ecc_vs_inc}, representing a composition of 45 million orbits.\footnote{Note that for the inclinations not shown ($0-40^\circ$, and the retrograde counterpart), only regular behavior was detected.} As expected from \citet{jD16a}, the resonances, appearing as thin lines in this phase-space magnification, are quite narrow and well separated at three Earth radii, but they widen and begin to interact as the semi-major axis is increased. At five Earth radii, the resonances stemming from the critical inclinations overlap and the phase space becomes divided into stable and unstable zones. Going beyond the regime of the navigation satellites (4-5 Earth radii), the resonant interactions become increasingly more pronounced, and, when near GEO ($\sim 6.7 R$), the stable band of polar orbits completely disappears and the prograde and retrograde regions become chaotically connected \citep[possibly allowing for chaotic flips in orientation from prograde to retrograde and back; see, e.g.,][]{vK94}. It is interesting to note that despite the classical result in the Hill problem \citep[q.v.][]{yS08}, we see here that retrograde orbits are not intrinsically more stable than their prograde counterparts.

\begin{figure}[htp!]
	\centering
	\includegraphics[width=0.55\textwidth]{cbarAVGFLIhor.pdf}
	\includegraphics[width=\textwidth]{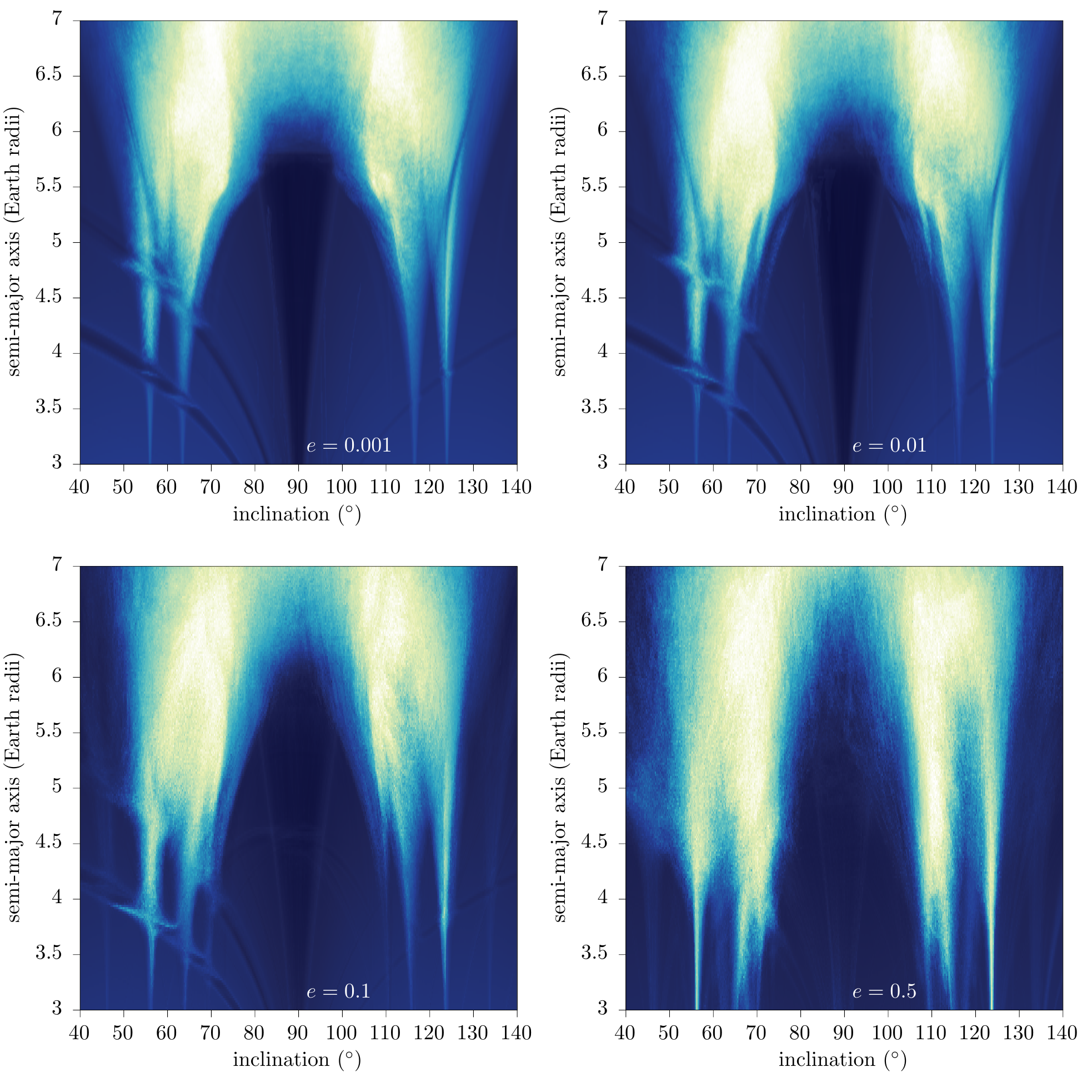}
	\caption{Dynamical stability maps in the inclination--semi-major axis phase space produced using the angle-averaged normalized FLI for increasing values of the orbit eccentricity. The colorbar for the FLI maps is normalized to 1.}
	\label{fig:sma_vs_inc_fli}
\end{figure}

And still another surprise, made more evident by Figure~\ref{fig:sma_vs_inc_fli}, is that quasi-circular orbits can be every bit as chaotic as eccentric ones. Figure~\ref{fig:sma_vs_inc_fli} shows a different slice or visualization of the phase space, fixing for each map the eccentricity instead of the semi-major axis. These dynamical maps manifest a number of striking and mysterious features, which require further theoretical investigations. We can point out the two critical inclination resonances that are present throughout all panels: $2 \dot\omega + \dot\Omega \approx 0$ at $56\fdg 1$ ($123\fdg 9$) and $\dot\omega \approx 0$ at $63\fdg 4$ ($116 \fdg 6$), both of which are of great importance for the Earth's navigation constellations \citep[qq.v.][]{aR15a,aR15b,jD16a,jD16b}. The two short arcs that cut across the prograde phase space at low semi-major axes, somewhat disrupting the two main inclinations, are from nodal resonances that primarily affect the orbit inclinations. These nodal commensurabilities seem to be washed out for very eccentric orbits (bottom right panel), and while they theoretically exist also for the retrograde regime, they are not as prominent there.

To further understand the evolution in the inclination--semi-major axis plane, we employ a different indicator based on a more practical quantity of interest; namely, the maximum eccentricity reached by an orbit during its evolution. The eccentricity growth has been used in the past for various astronomical applications to characterize the dynamical behavior of an orbit \citep[e.g.,][]{bKsD11,xR15}. In order to quantify the eccentricity growth through a wide range of semi-major axes and initial eccentricities we introduce a normalized eccentricity variation:
\begin{equation}
	\Delta e = \frac{| e_0 - e_{\text{max}}|}{|e_0 - e_{\text{re-entry}}|},
\end{equation}
where $e_0$ is the initial eccentricity, $e_{\text{max}}$ is the maximum eccentricity obtained during the propagation, and $e_{\text{re-entry}}$ is the value of the eccentricity that leads to re-entry for a given semi-major axis, as defined by our previous threshold (i.e., $a (1- e) < R + 120$ km). Regardless of the initial eccentricity, the quantity $\Delta e$ tends to 0 if the eccentricity is bounded around its initial value or to 1 if the eccentricity grows enough to reach the re-entry limit.

\begin{figure}[htp!]
	\centering
	\includegraphics[width=0.6\textwidth]{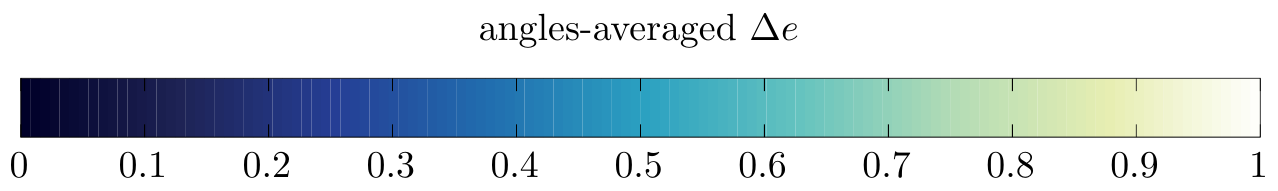}
	\includegraphics[width=\textwidth]{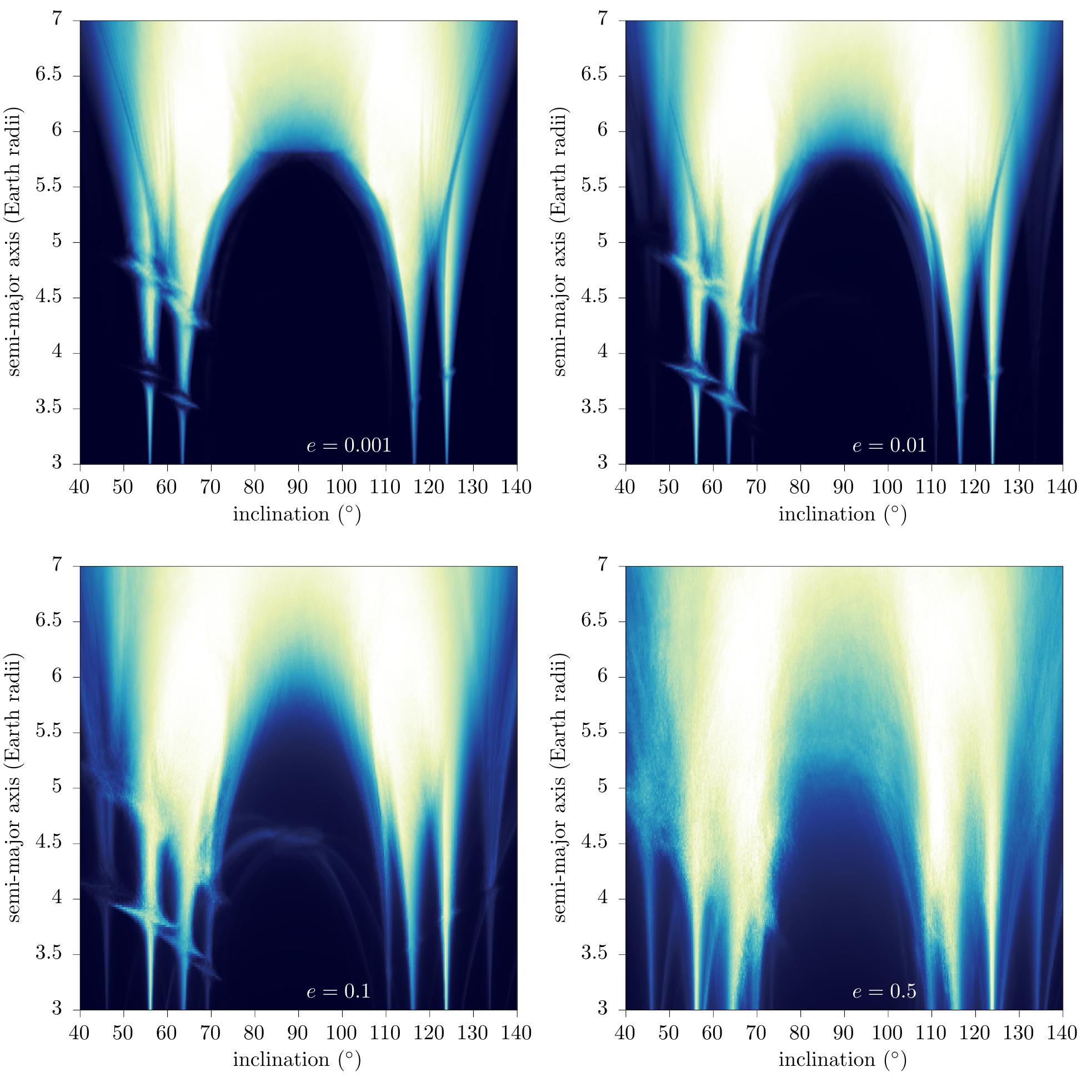}
	\caption{The inclination--semi-major axis phase space explored by means of the averaged normalized eccentricity variation $\Delta e$ for increasing values of initial eccentricity.}
	\label{fig:sma_vs_inc_delte}
\end{figure}

In Figure~\ref{fig:sma_vs_inc_delte} we redo the analysis of Figure~\ref{fig:sma_vs_inc_fli}, but using instead $\Delta e$ as a pseudo-chaotic indicator. Again, we produce an ensemble of 50 $\Delta e$ maps for randomly chosen initial phase sets and we report the average. The results for the four different values of initial eccentricity are in excellent agreement with those of Figure~\ref{fig:sma_vs_inc_fli} implying a strong connection between the local hyperbolicity and the transport of the eccentricity to large values, as previously suggested by \citet{jD16b}. On the other hand, the $\Delta e$ maps seem only to emphasize the chaotic features: resonant stable regions, like those connected with the nodal resonance in Figure~\ref{fig:sma_vs_inc_delte} are absent and can only be partially inferred from their interactions with the chaotic zones. These maps do, however, reveal an interesting feature for the $e = 0.1$ case (bottom left), showing lower altitude merging arches of the prograde and retrograde chaotic zones. A similar structure is also present in the FLI map, but is much fainter and more difficult to detect. 

\begin{figure}[htp!]
	\centering
	\includegraphics[width=0.6\textwidth]{cbarMAXECChor.pdf}
	\includegraphics[width=\textwidth]{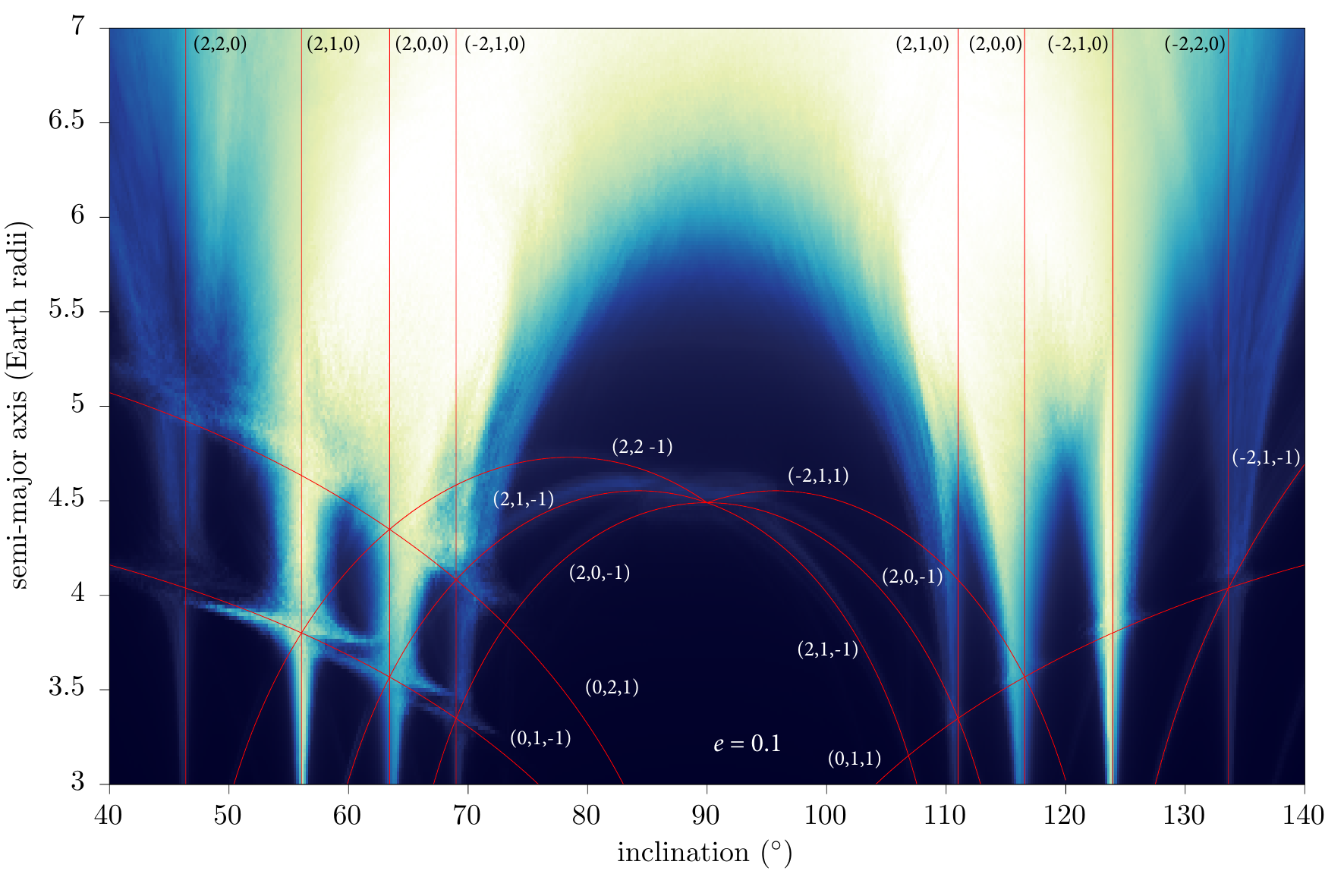}
	\caption{The averaged normalized eccentricity variation $\Delta e$ map for $e=0.1$, with the relevant lunisolar secular resonances (red lines) identified. The resonances are defined by $n_1 \dot\omega + n_2 \dot\Omega + n_3 \dot\Omega_2 = 0$, for integer coefficients $\ n_1 = \big\{ {-2}, 0 , 2 \big\}$, $n_2 = \big\{ 0, 1, 2 \big\}$, $n_3 \in \llbracket {-2}, 2 \rrbracket$ (not all zero), where only the effects of the $J_2$ perturbation on $\omega$ and $\Omega$ have been considered. (The symmetries between the prograde and retrograde resonances can be easily seen from Equations~\ref{eq:precess_rates}: when $n_2 = 0$, the resonances are symmetric about $i = 90^\circ$; when $n_2 \neq 0$, the prograde resonance $(n_1, n_2, n_3)$ is symmetric about $i = 90^\circ$ with the retrograde resonance $(-n_1, n_2, -n_3)$.)}
	\label{fig:sma_vs_inc_delte_resonances}
\end{figure}

We can pin down the dynamical origin of many of the features seen in the inclination--semi-major axis plane using the simple methodology of \citet{tE97} and \citet{aR15a}, to which we refer for more details. The observed features are the result of lunisolar secular resonances involving specific linear combinations of the averaged precession rates of the angles $\omega$, $\Omega$, and $\Omega_2$, approximately given by\footnote{We neglect here the lunisolar perturbations on the frequencies of nodal and apsidal precession, which is a valid approximation only at lower semi-major axes ($\lesssim 5 R$). At geosynchronous altitude, however, the lunisolar perturbations are of the same order as the secular oblateness term, and must be included to accurately map out the resonance web.} 
\begin{equation}
\label{eq:precess_rates}
\begin{array}{l}
	\ds \dot\omega = \frac{\omega_0}{2} \frac{5 \cos^2 i - 1}{(1 - e^2)^2}, \\[1.25em]
	\ds \dot\Omega = -\omega_0 \frac{\cos i}{(1 - e^2)^2}, \\[1.25em]
	\ds \dot\Omega_2 = -0.053\, \text{deg/d}.
\end{array}
\end{equation}
In Figure~\ref{fig:sma_vs_inc_delte_resonances}, the positions of all relevant secular resonances of the form $n_1 \dot\omega + n_2 \dot\Omega + n_3 \dot\Omega_2 = 0$, for integer coefficients $\ n_1 = \big\{ {-2}, 0 , 2 \big\}$, $n_2 = \big\{ 0, 1, 2 \big\}$, $n_3 \in \llbracket {-2}, 2 \rrbracket$ (not all zero), are superimposed on the $i-a$ averaged normalized eccentricity variation map for $e=0.1$. The main unstable structures emanate in a ``V'' shape at the birth of the critical inclination resonances $(n_1, n_2, 0)$. These `inclination-dependent-only' resonances widen and overlap with each other with increasing semi-major axes \citep[see Figure~14 in][]{jD16a}, rendering the highly inclined GEO region particularly unstable. In addition, their interaction with the nodal resonances $(0, n_2, n_3)$ creates very interesting structures in both the prograde and retrograde regions. Finally, the lower altitude merging arches can be identified with resonances of the form $(\pm 2, n_2, \mp 1)$.

\begin{figure}[htp!]
	\centering
	\includegraphics[width=0.45\textwidth]{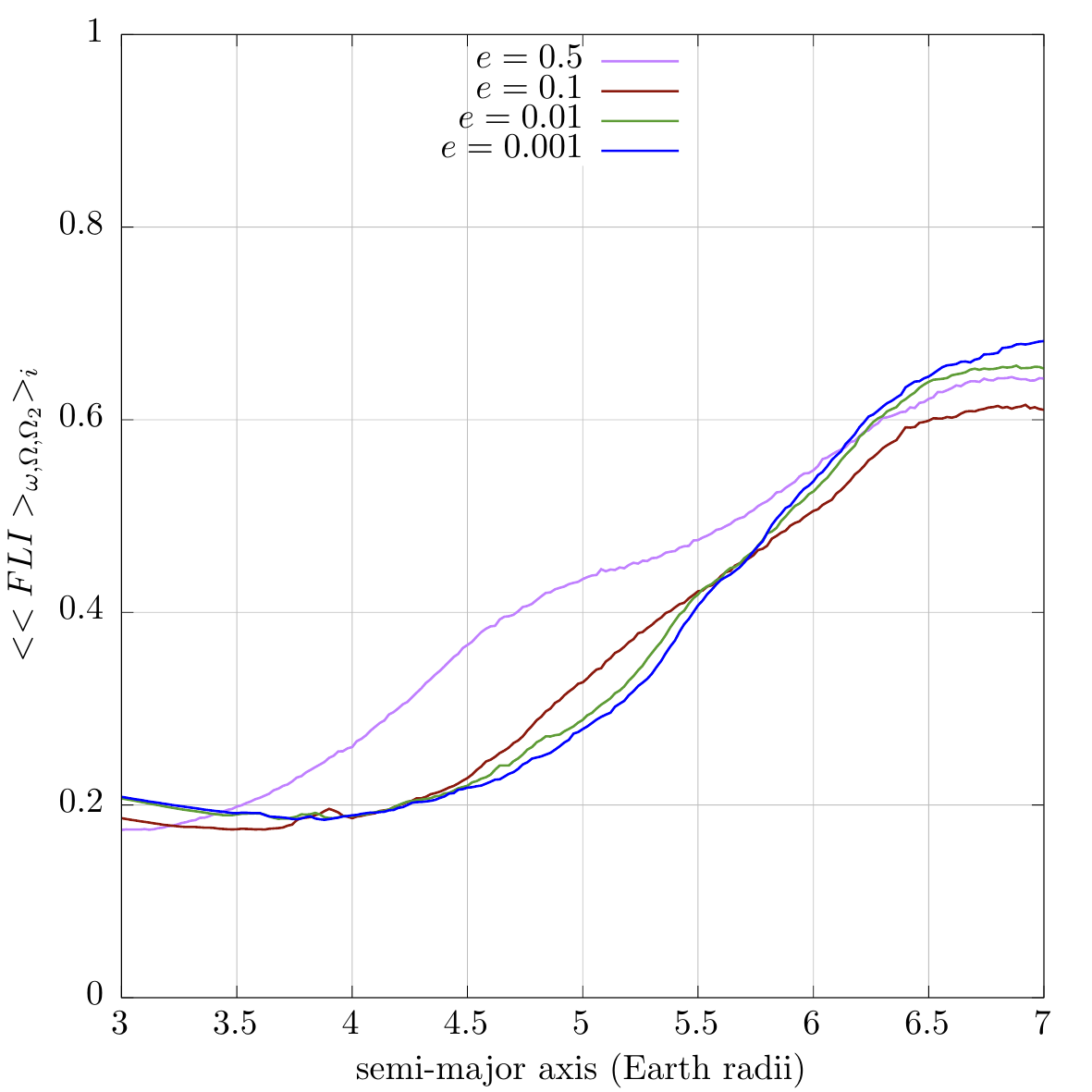}
	\includegraphics[width=0.45\textwidth]{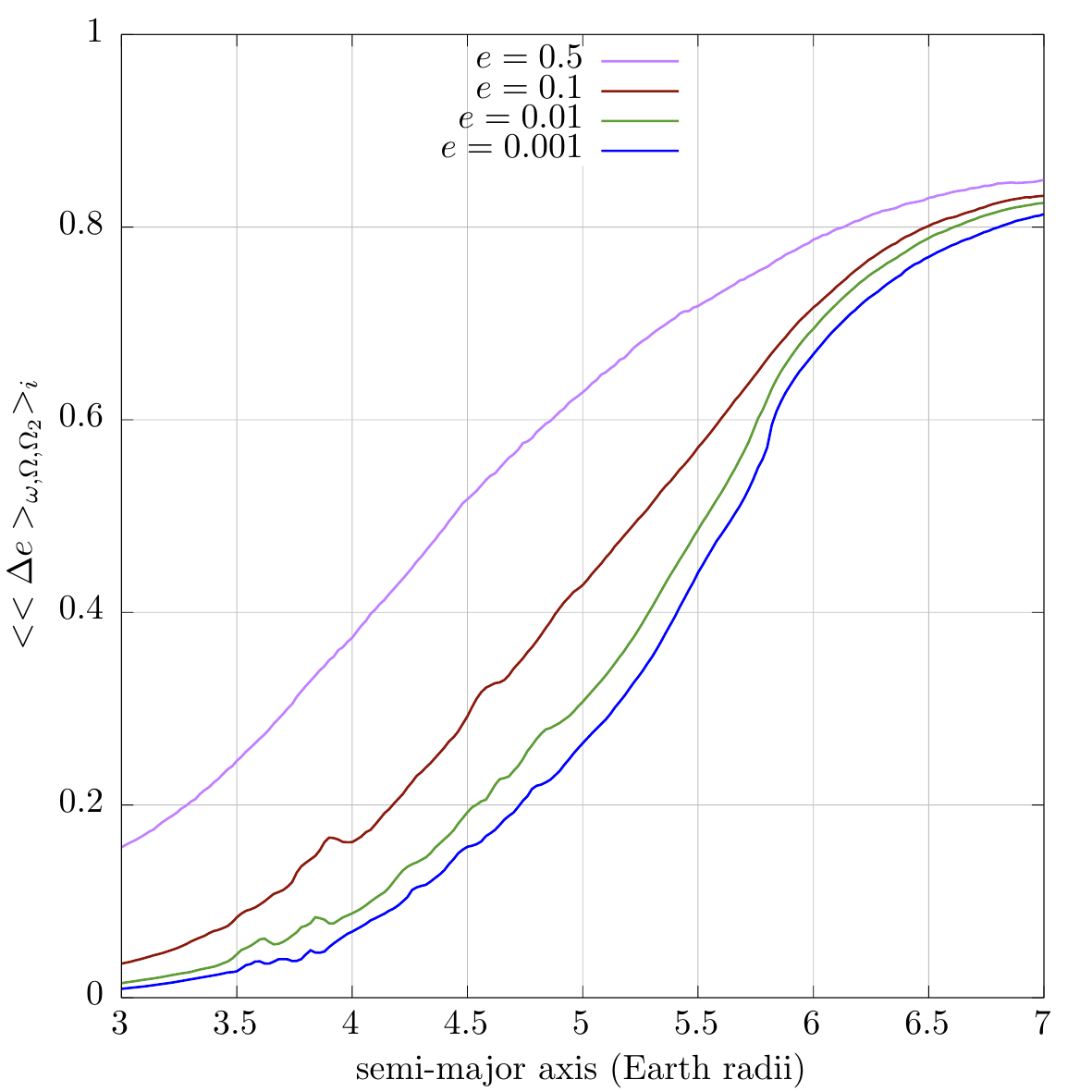}	
	\caption{A quantification of the transition from regular to chaotic behavior using the values of the FLI (left) and $\Delta e$ (right) indicators, averaged over the inclinations.}
	\label{fig:order_to_chaos}
\end{figure}

Figures~\ref{fig:sma_vs_inc_fli} and \ref{fig:sma_vs_inc_delte} have given us a clear picture of the complex behavior in the system as it undergoes a transition to chaos. In order to quantify this order-to-chaos transition, we compute the average value of the indicators over all inclinations for a fixed semi-major axis, as shown in Figure~\ref{fig:order_to_chaos} for the four values of initial eccentricity. The FLI curves (left panel) have very similar, nearly monotonically increasing behaviors across the different eccentricities, starting with an average FLI value of 0.2 at three Earth radii. This value stays constant up to about four Earth radii, and grows in a linear fashion up to six Earth radii, where it then stabilizes around 0.6. The more eccentric case, $e = 0.5$, has a more abrupt evolution, of course, but its initial and final state are the same as the near-circular cases. The $\Delta e$ curves (right panel) also have monotonically increasing evolutions, but with much larger slope, showing that the average eccentricity variations can reach in the order of 80 percent of the maximum possible at seven Earth radii. Clearly, highly elliptic orbits are the end state of most chaotic orbits at large geocentric distances.

\section{DISCUSSION}

It is natural to wonder how far in semi-major axis the chaotic Chirikov-like regime of resonance overlap will proceed. While a precise investigation must be carried out under octupole-order secular interactions, we speculate that far beyond the Laplace radius---the geocentric distance where the perturbing effects of oblateness and lunisolar forces are equal ($\sim 7.7$ Earth radii)---the critical inclination resonances that arise from the nodal and apsidal precession produced by Earth's oblateness will become less important. Hence, such distant orbits will likely be governed primarily by octupolar perturbations to the Lidov--Kozai dynamics \citep{sN16}, among other yet unstudied interactions with the lunar orbital precession frequencies. We should also point out that, strictly speaking, the classical Lidov--Kozai mechanism, in which the argument of perigee becomes stationary or librates allowing orbits to periodically exchange their eccentricity with inclination, simply does not exist in the Earth satellite problem. Under the entire domain of validity of the quadrupolar approximation, from which this basic resonance arises, the large oblateness of the Earth's figure completely destroys the Lidov--Kozai effect. Thus, the dramatic crash of the very distant, highly eccentric and highly inclined Soviet satellite \textit{Luna 3} (1959), was not likely caused by these types of dynamics, despite all folklore to the contrary \citep[see, for instance,][]{aM02}. In fact, the cause of \textit{Luna 3}'s decay remains a mysterious and intriguing problem for future investigation. 

It is rather ironic that the Earth's navigation satellites, whose current orbits are known to the utmost precision for the principle purpose of allowing for accurate positioning, navigation and timing \citep{tE97}, are inherently unpredictable.\footnote{Although we should note that the chaoticity only limits predictability over long time spans, it could actually help support greater insight over shorter timescales, given the high sensitivity of the trajectory to epoch states \citep{dS01}.} This dynamical situation, however, can be judiciously exploited for space debris remediation in the new paradigm of self removal of satellites through chaos and the associated orbital instabilities, from their otherwise long lifetime orbits \citep{aR15b}. Indeed, even in the GEO, where it has traditionally been thought that space objects would remain indefinitely, there exist many dynamical pathways that can be used to effectively clear this unique region of space from any permanent collision hazard \citep{vM14,cZ15}. It would be interesting, in this respect, to see whether increasing the satellite's area-to-mass ratio using a solar sail would promote the deorbiting process, through coupled gravitational and non-gravitational perturbations \citep{aRdS13} or by steering it into a short-lived resonance. 

These considerations emphasize the importance of investigating the analytical character of the resonances that drive these complicated dynamical behaviors and of developing accurate models that describe the resonant interactions \citep{jD16a}. We adopt here, perhaps, the simplest dynamical model that can capture the key physics. Our secular model of quadrupolar gravitational perturbations, which fundamentally traces back to Laplace in his study of the motion of natural satellites, is 2.5 degrees of freedom on account of the Moon's perturbed motion. Only one aspect of the Moon's non-Keplerian motion, namely, the regression of the line of nodes, is at the heart of the orbital instabilities observed in our atlas of FLI maps \citep{aR15a,jD16a,aCetal,aC16}. It must be underlined that our model is less accurate for the more distant orbits treated here, because terms depending upon the position of the Moon's perigee will appear in the octupole-order secular interactions \citep{pM61,rG75}. Thus, our model inherently neglects the interaction of the higher-order resonances associated with the harmonics in the octupole-level Hamiltonian. Clearly, these interesting distant orbits warrant further research, to cast light on the speculations made here.

The rich dynamics described in this paper are likely to be relevant for the problem of spacecraft trajectory design and navigation in binary asteroid systems \citep[q.v.][]{dS12}, which constitute an estimated 15\% of the near-Earth asteroids, or in the modeling of the evolution of the asteroid systems themselves. Our formalism can also be used to find longer-life trajectories suitable for science orbits around planetary satellites. 

Direct applications of these results to the natural satellite systems in the Solar System are rather few, at least in so far as their present configurations are concerned. Of the principal planetary satellites, only Saturn's satellite Iapetus orbits in a region controlled by the competing interior quadrupole moment of the planet and the external quadrupole potential from the Sun \citep{dBgC61,sT09}; but its inclination is nowhere near the critical values where chaos and instability set in. It is interesting to wonder how such dynamics would manifest in the high-obliquity ($\sim98^\circ$) Uranian system \citep{rM89,rF15} or in the Neptunian system, where the highly inclined ($\sim157^\circ$), retrograde orbit of Triton precesses with a period of about 680 years. The hypothetical polar rings around Neptune, conjectured by \citet{aD89}, would likely be unstable at the larger planetocentric distances where the instabilities stemming from the critical inclinations coalescence, although the approximation of Triton's perturbing potential by its quadrupole component is surely poor, as \citet{sT09} note.  

The dynamics described in this paper, arising from a simple time-periodically driven Hamiltonian system with two degrees of freedom, could play a role in exoplanetary systems \citep{sD14}. A warm Jupiter---a jovian-size planet in a relatively temperate Mercury-like orbit---flanked by an outer companion in an eccentric and mutually inclined orbit and an inner hot Jupiter (augmenting the quadrupole moment of the host star) could experience similar destabilizing quadrupole-order effects. Perhaps similar dynamical quadrupole interactions could account for the apparent lack of additional, nearby planetary companions in hot-Jupiter systems \citep{cH16} or provide other architectural insights. One, therefore, does not need to invoke octupole-order secular interactions \citep[qq.v.][]{bL15,sN16} to find large eccentricity excitations or  chaotic flips in orientation \citep[see][for such bizarre effects in the seemingly more mundane world of Earth satellites]{vK94}. 

\section{CONCLUSION}

The number of artificial Earth satellites is now at least an order of magnitude larger than their natural counterparts in the Solar System, and the number of space debris, brought on by the Space Age, is much larger still. It is the structure of the satellite and the nature of its orbit that determines which perturbations are significant and which are negligible. In this sense, every distinct problem conditions its own particular scheme of computation, and many refinements, sometimes reducing the always elaborate calculations in a marked degree, depend on a careful examination of the dynamical situation. We can gain a better understanding by analyzing the simplest possible cases and by leaving out of our first attempts all intricate complications. One of the main complications that was stripped away here was the octupole-order secular interactions. As a result, this paper was chiefly concerned with the role played by the lunar nodal precession, the source of the Hamiltonian's explicit time dependence, in a quadrupolar model of oblateness and lunisolar perturbations. We showed that the simple and deterministic equations of \citet{sT14}, when accounting for the precession of the Moon's orbital axis, can exhibit a variety of interesting and hitherto unknown dynamical behaviors, answering why, even without the destabilizing influence of atmospheric drag, many of Earth's satellites eventually fall down. Further studies in this area may lead to deeper insights in celestial mechanics as well as provide practical results for satellite technology.

\begin{acknowledgements}
The title of this paper is derived from and pays homage to \citet{cF00}. 
A deep debt of gratitude is extended to the AUTH Compute Infrastructure and Resource, where the numerical simulations were mainly hosted, and for the support provided by Paschalis Korosoglou and the AUTH Scientific Computing Office.  We are grateful to the anonymous referee for many valuable suggestions.
This paper has also benefited from useful comments made by C. Efthymiopoulos, N. Todorovi\'{c}, and K. Tsiganis. 
JD thanks M.\,Fouchard for continued discussions about the FLI and P.\,Chainais for the evocation of its possible modification. 
AR wishes to thank D.\,Amato, M.\,Jankovi\'{c}, and M.\,Vetrisano, of the Stardust Network, for many stimulating fisica discussions.
This work is partially funded by the European CommissionÕs Framework Programme 7, through the Stardust Marie Curie Initial Training Network, FP7-PEOPLE-2012-ITN, Grant Agreement 317185. JD acknowledges the partial financial support from the Australian Research Council project (ID LP130100243). Our research has made use of NASA's Astrophysics Data System.
\end{acknowledgements}

\appendix

\section{Model Validation} \label{sec:validate}

Numerical integration of high-fidelity dynamical models represents today the most accurate means of calculating the exact trajectory of a space object in a given time interval. While such computations can give ``precise'' solutions for specific initial conditions, these afford little insight into the nature of the problem or the essential dependence of the perturbed motion on the system parameters. The efficient use of the computer, moreover, depends in a highly important way on an understanding of these dependencies, and on a careful preliminary selection of the force models to be integrated. Furthermore, such numerical propagation, taking into account both short-period and long-period terms, becomes very costly when continuously applied over hundreds of thousands of revolutions or more, as is necessary, for example, when studying the long-term stability of orbits. 

Our purpose here is to adopt the simplest possible expressions useful for studying the long-term evolution of Earth satellite orbits in the region of semi-major axes between three and seven Earth radii. These expressions must reveal the basic qualitative regularities of motion, and they must provide, with a certain degree of accuracy, quantitative predictions of the long-term changes. Averaging methods have proved to be among the most powerful instruments for the study of dynamical systems. \citet{jD16a} computed the FLI stability maps for the MEO problem using a hierarchy of more realistic singly averaged models, and found that the basic model of quadrupolar gravitational interactions of oblateness and lunisolar perturbations was sufficient to capture nearly all of the qualitative and quantitative features (dynamical structures, degree of hyperbolicity, domains of collision orbits, etc.) of the more complicated models.\footnote{These other models included, among other things, the oft-assumed important tesseral resonant effects, which can produce small, localized instabilities in the satellite's semi-major axis \citep{tE97}.} Thus, the comparison of FLI maps produced from the basic physical model of \citet{jD16a} with the results obtained from our doubly averaged formulae (Section~\ref{sec:Milan}) is a significantly reliable estimate of the accuracy of the approximated equations. Such comparison permits us to conclude about the applicability of the doubly averaged formulation for considering the evolution of actual satellite orbits.    

The model of \citet{jD16a}, a Fortran prototype of the software, STELA,\footnote{STELA (Semi-analytic Tool for End of Life Analysis) can be downloaded from the CNES website:\\ https://logiciels.cnes.fr/content/stela} is a singly averaged formulation of perturbed Keplerian motion, written in the equinoctial variables, and uses a precise ephemeris for the Moon and the Sun. The simpler model used here, detailed in Section~\ref{sec:Milan}, on the other hand, is a doubly averaged, vectorial formalism, that takes a  Keplerian ellipse for the apparent solar motion and a precessing ellipse for that of the Moon. Both models only account for quadrupole-order secular interactions.   

\begin{figure}[t!]
	\centering
	\includegraphics[width=\textwidth]{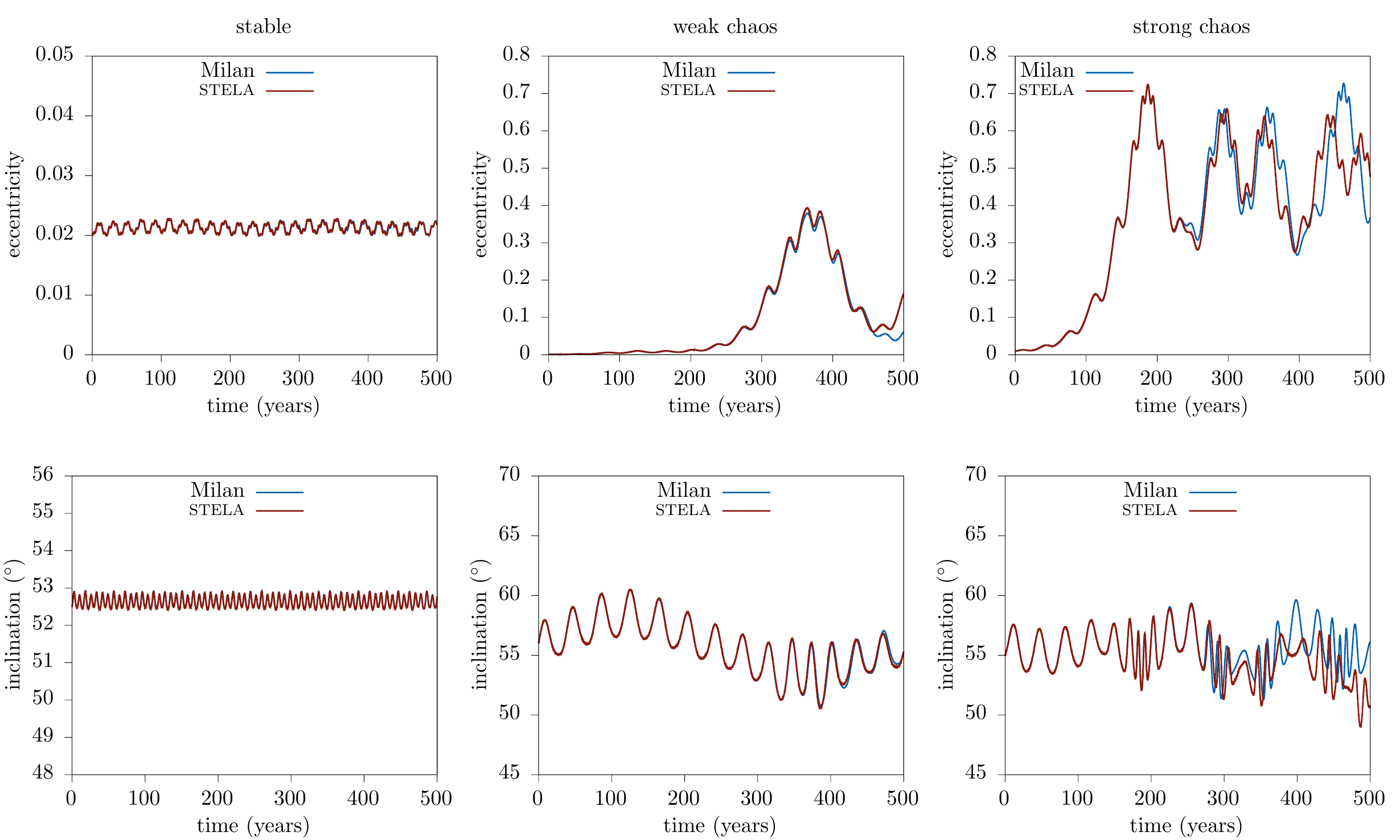}	
	\caption{The eccentricity (top) and inclination (bottom) evolution of three MEO trajectories with distinct dynamical character: stable (left), weakly chaotic (middle), and strongly chaotic (right). The orbits are propagated with STELA (red lines) and our `Milan' model (blue lines).}
	\label{fig:orbit_compare}
\end{figure}

\begin{figure}[htp!]
	\centering
	\includegraphics[width=0.6\textwidth]{cbarFLIhor.pdf}
	\includegraphics[width=\textwidth]{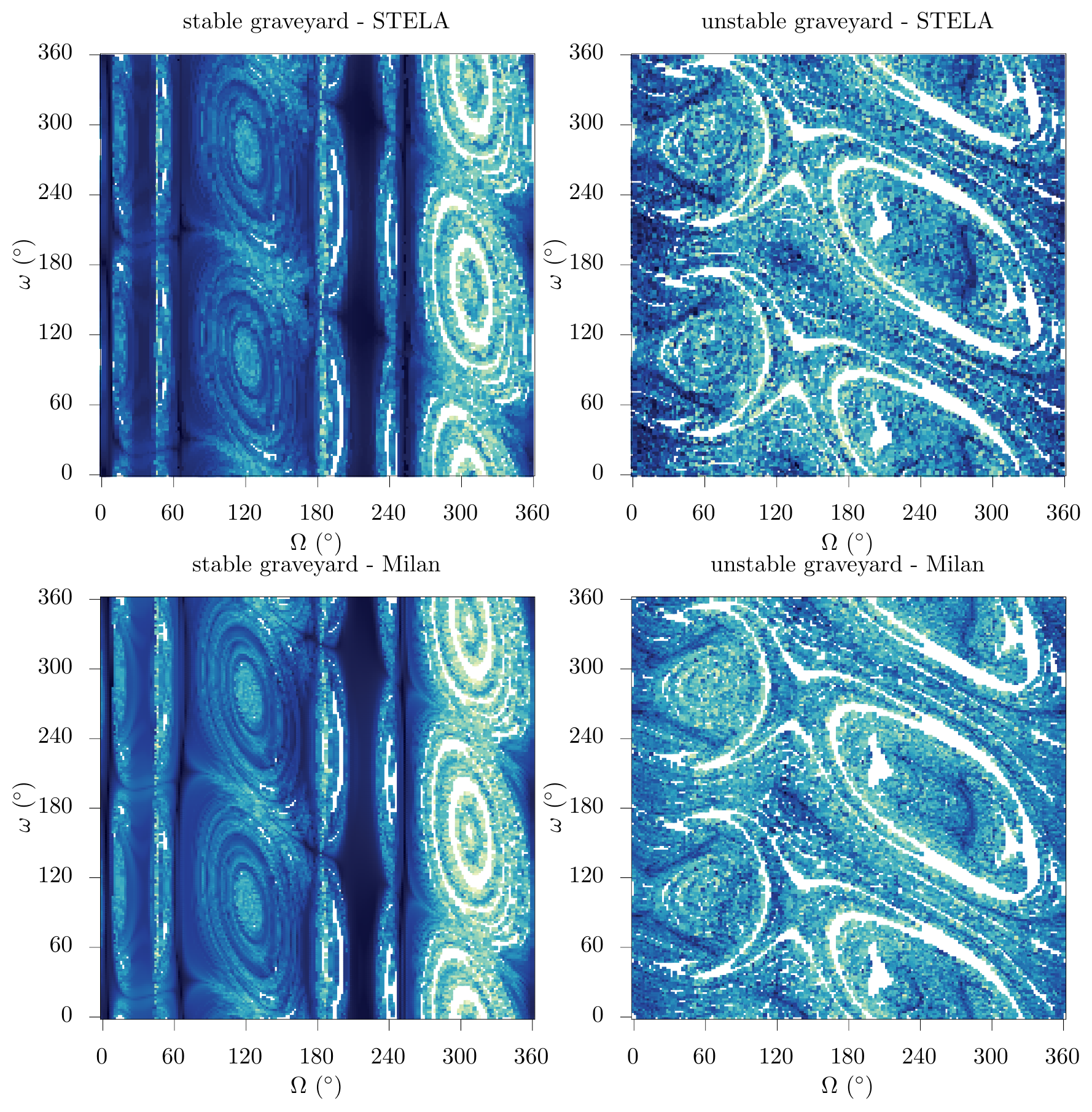}	
	\caption{The FLI stability maps, computed using STELA (top), for a study of stable (left) and unstable (right) graveyard scenarios for the Galileo disposal strategy, as detailed in \citet{aR15b}. The maps produced by Milan (bottom) are in excellent agreement.}
	\label{fig:fli_compare}
\end{figure}

Figure~\ref{fig:orbit_compare} shows the evolution of three trajectories with distinct dynamical behaviors: a stable orbit (left), a weakly chaotic orbit (middle), and a strongly chaotic orbit (right). Each orbit is propagated with both STELA and our simple `Milan' model. The time evolution of the eccentricity and the inclination are identical in the case of the stable orbit, but eventually diverge for the chaotic orbits (as expected), at approximately 450 years for the weakly chaotic orbit and at approximately 250 years for the strongly chaotic orbit. These timescales are associated with the Lyapunov time of each orbit, i.e., the average time after which nearby orbits diverge exponentially. However, as far as the dynamical characterization of the orbits is concerned, the two models give similar values of the FLI, hence, for the purposes of stability analysis, both models should lead to the same dynamical taxonomy. Figure~\ref{fig:fli_compare} shows a comparison of the FLI stability maps in the $\Omega-\omega$ plane, for fixed actions ($a$, $e$, and $i$), for two proposed Galileo graveyard orbits, taken from \citet{aR15b}, to which we refer for the omitted details and discussion. The maps produced from the doubly averaged model match incredibly well with those produced from the more complicated STELA model. Not only does Milan capture in a convincing fashion all the stable domains in the maps, but also reproduces in a very accurate manner the structure and position of regions of initial conditions that are inherently chaotic and that lead to re-entry.

\section{An Averaged FLI} \label{sec:FLI}

Regardless of whether they are variational or spectral, numerical cartographic techniques of the phase space are essential for an orbit's stability taxonomy \citep{jL92,cF00,cS10}. Specifically in Celestial Mechanics, the computation of numerical indicators over grids of initial conditions is one of the major tools to study the long-term stability of a given dynamical system. The variational methods, among which we find the FLI, are related to the evolution of the tangent vector $w$ with the passage of time. We recall that given a first-order $n$-dimensional autonomous system of ordinary differential equations,
\begin{align}
	\dot{x}=f(x), \ x=(x_{1},\cdots,x_{n}),
\end{align}
by integrating the variational equation 
\begin{align}
	\dot{w}=\Big(\frac{\partial f}{\partial x}\Big)w, \ w=(w_{1},\cdots,w_{n}),
\end{align}
the FLI at time $t$, defined as 
\begin{align}
	\textrm{FLI}(t;x_{0},w_{0}) = \sup_{\tau \le t} \log \vert\vert w(\tau) \vert\vert
\end{align}
where $\vert \vert \bullet \vert \vert$ denotes the Euclidian $L_{2}$-norm, is able to distinguish (in a relatively short CPU time) between ordered, resonant, and chaotic motions (a given initial condition $(x_{0},w_{0})$ being specified). 

The value of FLI depends basically on the initial state vector $x_{0}$ and integration time, since the choice of the initial deviation vector $w_{0}$ does not affect the results (Note that because particular orientations of the initial deviation vector might lead to spurious structures, therefore a normalized randomly oriented one was used here.). In perturbed Keplerian motion, like that studied here, the state is often defined as a set of elements that characterize the instantaneous orbit. There are three action-like elements ($a,e,i$) and three corresponding angles ($\omega,\Omega,M$). In our averaged formulation, however, only four of those elements vary ($e,i,\omega,\Omega$), while $a$ is constant and treated as a parameter, and there is a time dependence stemming from the precessing lunar orbit ($\Omega_2$). 

In cartographic studies of the asteroid belt, various techniques have been used to study its global dynamics. Far from resonances, any choice of the initial angles would lead to similar results. Thus, a zero angle initial state vector could be used \citep[see, e.g.,][]{pR01}. On the other hand, the regions of phase space where a single resonance acts can be studied with an elegant pendulum-like modeling. For the numerical investigation of these regions, we can choose two sets of initial angles: one in which their resonant combination corresponds to the stable equilibrium of the pendulum and one in which their resonant combination corresponds to the unstable equilibrium. Those two maps contain all the useful information for this region.

Unfortunately, neither of the above is the case in our study. The web of lunisolar secular resonances in Earth satellite orbits is so dense \citep{tE97,aR15a} that a different approach is required. \citet{nTbN15} proposed the use of a random selection of initial angles at each point of the actions' grid to reveal the full width of the chaotic regions. Here, we take this idea one step further. Instead of randomly selecting the triplets ($\omega,\Omega,\Omega_2$) at each grid point, we fix a randomly selected triplet ($\omega,\Omega,\Omega_2$) for which we compute a whole FLI dynamical map. After a proper normalization, we repeat this computation for other randomly selected triplets and we finally average the normalized FLI values at each point in the actions' grid.

\begin{figure}[ht!]
	\centering
	\includegraphics[width=0.575\textwidth]{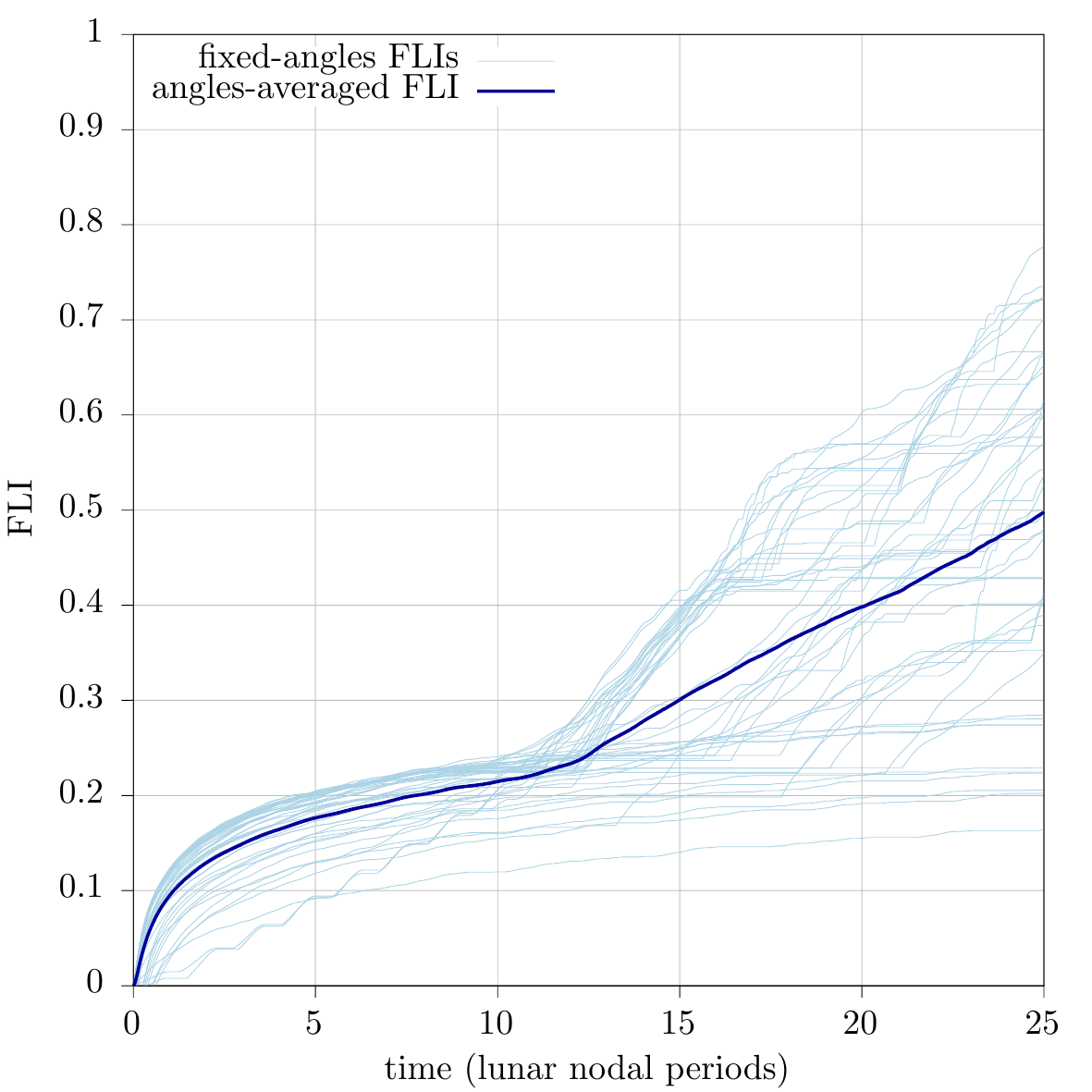}	
	\caption{Time evolution of the FLI for 50 random initial phase sets for a given ($a,e,i$) and the averaged normalized FLI of this ensemble, showing how the dynamical taxonomy of a fixed action changes based on the initial angles.}
	\label{fig:avgFLI_timeSeries}
\end{figure}

In Figure~\ref{fig:avgFLI_timeSeries} we present the time evolution of the angle-averaged normalized FLI over an ensemble of orbits for a specific set of ($a,e,i$) and 50 randomly selected values of the triplet ($\omega,\Omega,\Omega_2$). The value reported in our FLI maps in Figures~\ref{fig:ecc_vs_inc} and \ref{fig:sma_vs_inc_fli} is the value of the averaged FLI at the end of the 465 year integration span.

More formally, to cut loose the angle dependencies in the stability maps, taking advantage of the reduced (and justified, cf. Appendix \ref{sec:validate}) physics implemented in our equations of motion, we have introduced an angle-averaged FLI indicator, with the averaging being performed spatially. 
For the sake of simplicity, we suppose that the dynamics may be written in terms of  \textit{slow-fast} components and we decompose the vector $x$ as $x=(I,\phi)$, where $I_{1},\dots,I_{m} \in \mathbb{R}$ are the slow variables and $\phi_{1},\dots,\phi_{p} \in \mathbb{S}$ the fast variables.\footnote{We consider here the general \textit{standard system} framework to suggest the important Hamiltonian case in which we perturb an integrable system. In that case $p=m$ and the variables $(I,\phi)$ are then the action-angle variables.}  
Then, for a selected action $I_{0}$ (an initial seed), the averaged FLI reads as
\begin{equation}
	\label{eq:wFLI}
	\overline{\textrm{FLI}}_{\Gamma}(t;I_{0},w_{0}) 
	\equiv \Big\langle \textrm{FLI}(t;I_{0},\phi,w_{0}) \Big\rangle_{\Gamma},
\end{equation}
where $\Big\langle \bullet \Big\rangle_{\Gamma}$ denotes the (angle) space average over the domain $\Gamma$. For simplicity, we will simply denote this now as $\overline{\textrm{FLI}}(I_{0})$.

In our case study, this averaging has been performed over the three angles of the system, namely, $\omega$, $\Omega$, and $\Omega_{2}$. For a given domain of integration $\Gamma \subset  [0,\pi] \times \mathbb{T}^{2}$, we have approximated the space average in Equation~\eqref{eq:wFLI} via its  discrete  counterpart, i.e. by using the quantity   
\begin{equation}
	\frac{1}{N} \sum_{i=1}^{N} \textrm{FLI}(t;I_{0},\phi_{i},w_{0}), \ N \gg 1, \{\phi_{i}\}_{i=1}^{N} \in \Gamma 	\subset [0,\pi] \times \mathbb{T}^{2}.
\end{equation}
Therefore, this indicator requires that we compute, for a selected action $I_{0}$, $N$-FLIs. 

In this regard, the averaged indicator $\overline{\textrm{FLI}}$ can be interpreted as an effective FLI providing, asymptotically with large $N$, the mean degree of hyperbolicity of the system, independent of the angles, for a selected and fixed action $I_{0}$.\footnote{Depending on the application, refinements or variations of the definition of the averaged indicator (\ref{eq:wFLI}) are straightforward. For example, we can introduce a weighted version of the averaged version, i.e. we can replace (\ref{eq:wFLI}) by $\rho \big\langle \textrm{FLI}(t;I_{0},\phi,w_{0}) \big\rangle_{\Gamma}$ where $\rho \in \mathbb{R}$ is any coefficient defined via the set of the N-FLIs. The quantity $\rho$ may reflect by its nature (according to its definition) a kind of ``confidence-coefficient,'' or may either penalize large or small FLIs values, tending to take into consideration the dispersion of data, etc.}



\begin{thebibliography}{}

\bibitem[Allan \& Cook(1964)]{rAgC64}
Allan, R. R., \& Cook, G. E. 1964, 
RSPSA, 280, 97

\bibitem[Breiter(2001)]{sB01}
Breiter, S. 2001,
CeMDA, 81, 81

\bibitem[Brouwer \& Clemence(1961)]{dBgC61}
Brouwer, D., \& Clemence, G. M. 1961, 
Planets and Satellites, 
G. P. Kuiper \& B. M. Middlehurst, 
Chicago: University of Chicago Press, 31


\bibitem[Celletti et al.(2016a)]{aC16}
Celletti, A., Gale\c{s}, C., \& Pucacco, G. 2016,
SIAM J. Applied Dynamical Systems, 15, 1352

\bibitem[Celletti et al.(2016b)]{aCetal}
Celletti, A., Gale\c{s}, C., Pucacco, G., Rosengren, A. J. 2015, 
CeMDA, DOI:10.1007/s10569-016-9726-8

\bibitem[Cooper et al.(2015)]{nC15}
Cooper, N. J., Renner, S., Murray, C. D., \& Evans, M. W. 2015, 
\aj, 149, 27 

\bibitem[Daquin et al.(2016a)]{jD16a}
Daquin, J., Rosengren, A. J., Alessi, E. M., Deleflie, F., Valsecchi, G. B., \& Rossi, A. 2016,
CeMDA, 124, 335

\bibitem[Daquin et al.(2016b)]{jD16b}
Daquin, J., Rosengren, A. J., \& Tsiganis, K. 2016,
arXiv:1606.00106

\bibitem[Dobrovolskis et al.(1989)]{aD89}
Dobrovolskis, A. R., Borderies, N. J., \& Steiman-Cameron, T. Y. 1989, 
\icarus, 81, 132

\bibitem[Dong et al.(2014)]{sD14}
Dong, S., Katz, B., \& Socrates, A. 2014,
\apjl, 781, L5

\bibitem[Ely \& Howell(1997)]{tE97}
Ely, T. A., \& Howell, K. C. 1997, 
DSSys, 12, 243


\bibitem[French et al.(2015)]{rF15}
French, R. G., Dawson, R. I., \& Showalter, M. R. 2015,
\aj, 149, 142

\bibitem[Froeschl\'{e} et al.(2000)]{cF00}
Froeschl\'{e}, C., Guzzo, M., \& Lega, E. 2000,
Sci, 289, 2108

\bibitem[Greenberg(1975)]{rG75}
Greenberg, R. 1975,
\mnras, 170, 295


\bibitem[Hough(1981)]{mH81}
Hough, M. E. 1981, 
CeMec, 25, 111

\bibitem[Huang(2016)]{cH16}
Huang, C., Wu, Y., Triaud, A. H. M. J. 2016, 
arXiv: 1601.05095

\bibitem[Katz \& Dong(2011)]{bKsD11}
Katz, B., \& Dong, S. 2011,
arXiv:1105.3953

\bibitem[Kudielka(1994)]{vK94}
Kudielka, V. W. 1994,
CeMDA, 60, 455

\bibitem[Kudielka(1997)]{vK97}
Kudielka, V. W. 1997,
The Dynamical Behaviour of our Planetary System, 
R. Dvorak \& J. Henrard, 
Dordrecht: Kluwer Academic Publishers, 243 

\bibitem[Laskar et al.(1992)]{jL92}
Laskar, J., Froeschl\'{e}, C., \& Celletti, A. 1992, 
PhyD 56, 253

\bibitem[Lidov \& Yarskaya(1974)]{mLmY74}
Lidov, M. L., \& Yarskaya, M. V. 1974, 
KosIs, 12, 139 

\bibitem[Liu et al.(2015)]{bL15}
Liu, B., Mu\~{n}oz, D. J., \& Lai, D. 2015,
\mnras, 447, 747

\bibitem[Malhotra et al.(1989)]{rM89}
Malhotra, R., Fox, K., Murray, C. D., \& Nicholson, P. D. 1989,
\aap, 221, 348

\bibitem[Morand et al.(2014)]{vM14}
Morand, V., Fraysse, H., Lamy, A., Le Fevre, C., Pinede, R. 2014,
ISSFD, 24, S10-6

\bibitem[Morbidelli(2002)]{aM02}
Morbidelli, A. 2002, 
Modern Celestial Mechanics: Aspects of Solar System Dynamics
(London: Taylor \& Francis)

\bibitem[Musen(1961)]{pM61}
Musen, P. 1961,
JGR, 66, 2797

\bibitem[Naoz(2016)]{sN16}
Naoz, S. 2016, 
arXiv:1601.07175



\bibitem[Peale(1999)]{sP99}
Peale, S. J. 1999,
\araa, 37, 533

\bibitem[Ramos et al.(2015)]{xR15}
Ramos, X. S., Correa-Otto, J. A., \& Beaug\'{e}, C. 2015, 
CeMDA, 123, 453


\bibitem[Robutel \& Laskar(2001)]{pR01}
Robutel, P., \& Laskar, J. 2001,
\icarus, 152, 4

\bibitem[Rosengren \& Scheeres(2013)]{aRdS13}
Rosengren, A. J., \& Scheeres, D. J. 2013, 
AdSpR, 52, 1545

\bibitem[Rosengren et al.(2015)]{aR15a}
Rosengren, A. J., Alessi, E. M., Rossi, A., Valsecchi, G. B. 2015,
\mnras, 449, 3522

\bibitem[Rosengren et al.(2016)]{aR15b}
Rosengren, A. J., Daquin, J., Tsiganis, K., Alessi, E. M., Deleflie, F., Rossi, A., Valsecchi, G. B. 2016,
\mnras, DOI:10.1093/mnras/stw2459

\bibitem[Scheeres(2012)]{dS12}
Scheeres, D. J. 2012,
Orbital Motion in Strongly Perturbed Environments: 
Applications to Asteroid, Comet and Planetary Satellite Orbiters
(Berlin: Springer-Praxis)

\bibitem[Scheeres et al.(2001)]{dS01}
Scheeres, D. J., Han, D., \& Hou, Y. 2001, 
JGCD, 24, 573

\bibitem[Shen \& Tremaine(2008)]{yS08}
Shen, Y., \& Tremaine, S. 2008, 
\aj, 136, 2453

\bibitem[Skokos(2010)]{cS10}
Skokos, Ch. 2010, 
LNP, 790, 63

\bibitem[Struve(1928)]{gS28}
Struve, G. 1928,
\aj, 38, 193

\bibitem[Todorovi\'{c} and Novakovi\'{c}(2015)]{nTbN15}
Todorovi\'{c}, N., \& Novakovi\'{c}, B. 2015,
\mnras, 451, 1637

\bibitem[Tremaine et al.(2009)]{sT09} 
Tremaine, S., Touma, J., \& Namouni, F. 2009, 
\aj, 137, 3706 

\bibitem[Tremaine \& Yavetz(2014)]{sT14} 
Tremaine, S., \& Yavetz, T. 2014,
AmJPh, 82, 769

\bibitem[Upton et al.(1959)]{pM59}
Upton, E., Bailie, A., \& Musen, P. 1959, 
Sci, 130, 1710

\bibitem[Vashkov'yak(1974)]{mV74}
Vashkov'yak, M. A. 1974, 
KosIs, 12, 757

\bibitem[Zhao et al.(2015)]{cZ15}
Zhao, C. Y., Zhang, M. J., Wang, H. B., Xiong, J. N., Zhu, T. L., \& Zhang, W. 2015,
AdSpR, 56, 377

\end{thebibliography}
\end{document}